\documentclass[a4paper,11pt]{article}
\pdfoutput=1 

\usepackage{jcappub} 

\usepackage[T1]{fontenc} 
\usepackage{booktabs}
\usepackage{graphicx}
\usepackage{amsmath,amssymb}
\graphicspath{{./figures/}}
\usepackage{appendix}
\usepackage{placeins}
\usepackage[normalem]{ulem}
\usepackage{multirow}
\usepackage{bm}
\usepackage{times}
\usepackage{comment}
\usepackage{mathtools}
\usepackage{float}
\usepackage{mathabx}

\usepackage{amsfonts}
\usepackage{upgreek}
\usepackage{latexsym}
\usepackage{stfloats}
\usepackage{afterpage}
\usepackage{caption}
\usepackage{subcaption}
\usepackage{enumitem}
\usepackage{orcidlink}

\newcommand{\be}{\begin{equation}}
\newcommand{\ee}{\end{equation}}
\newcommand{\beq}{\begin{equation}}
\newcommand{\eeq}{\end{equation}}
\newcommand{\bea}{\begin{eqnarray}}
\newcommand{\eea}{\end{eqnarray}}

\newcommand{\UU}{{\cal U}}

\newcommand{\De}{\Delta}

\newcommand{\ga}{\gamma}

\newcommand{\de}{\delta}

\newcommand{\lam}{\lambda}

\newcommand{\si}{\tau}

\newcommand{\pa}{\parallel}

\newcommand{\A}{{\cal A}}
\newcommand{\BB}{{\cal B}}
\newcommand{\CC}{{\cal C}}

\newcommand{\obs}{{\rm o}}
\newcommand{\emi}{{\rm e}}

\renewcommand{\be}{\beta}

\newcommand{\bean}{\begin{eqnarray*}}
\newcommand{\eean}{\end{eqnarray*}}

\newcommand{\id}{{\rm 1\kern -2.5pt I}}

\title{\boldmath 
Complete Second-Order Relativistic Derivation of the Observed Pulsar Timing Modulations}

\author[a]{Matteo Magi\orcidlink{0000-0001-8062-9447}}

\author[b,c]{and Jaiyul Yoo\orcidlink{0000-0003-4988-8787}}

\affiliation[a]{Cosmology, Gravity and Astroparticle Physics Group, Center for Theoretical Physics of the Universe, Institute for Basic Science (IBS), Daejeon, 34126, Korea}

\affiliation[b]{Center for Theoretical Astrophysics and Cosmology, Department of Astrophysics,
University of Zurich, Winterthurerstrasse 190, CH-8057, Zurich, Switzerland}
\affiliation[c]{Department of Physics, University of Zurich, Winterthurerstrasse 190, CH-8057, Zurich, Switzerland}

\emailAdd{mmagi@ibs.re.kr}

\emailAdd{jaiyul.yoo@uzh.ch}

\abstract{
We present a fully nonlinear relativistic description of the observed pulsar timing modulations, defined through the ratio of the proper-time intervals between two successive emissions of radio pulses and their observations by a pulsar timing array. Working in the limit of a vanishing emission interval, we derive the observed pulsar timing modulation to second order in relativistic perturbation theory around a Minkowski background, without fixing a gauge and retaining the full scalar, vector, and tensor content of the metric perturbations. Although the derivation naturally introduces several gauge-dependent intermediate quantities, we explicitly demonstrate that they combine into a coordinate-independent expression. Moreover, we explicitly show that the second-order timing modulation from two neighboring radio pulses coincides with the observed redshift, defined as the fractional change in wavelength along a single geodesic. The agreement represents a nontrivial second-order realization of the exact nonlinear equivalence proven in 
Magi \& Yoo~2026 \cite{MAGIYOO26}.  
}

\begin{document}
\maketitle

\section{Introduction}
Pulsar timing arrays (PTAs) consist of ensembles of millisecond pulsars whose radio pulses are monitored over observational baselines of years to decades. The measured times of arrival are compared with timing models describing the pulsars' spin evolution and astrometric or binary motion, together with propagation and Solar-System effects, leaving timing residuals that may contain additional correlated signals \cite{IPTA10,PPTA13}. Gravitational waves perturb the propagation of the radio pulses and induce characteristic correlations among pulsars across the sky, with an isotropic stochastic background giving rise to the Hellings--Downs angular correlation \cite{HEDO83}. Their long observational baselines make PTAs sensitive to gravitational waves in the nanohertz band, complementary to the frequency ranges probed by ground- and space-based interferometers \cite{LIGO15,LISA24red}. Recent results from the major PTA collaborations have reported evidence for a common red signal with spatial correlations consistent with a gravitational-wave background \cite{NANO23gw,EPTA23C,PPTA23gw,CPTA23gw}.

The theoretical description of pulsar timing signals has traditionally been developed within linearized gravity, where metric perturbations induce a first-order modulation of the pulse arrival times \cite{ESWA75,DETWE79,ANBAET09}. At this order, the standard PTA response can be understood as the redshift of the radio signal along its null trajectory (i.e., the Sachs-Wolfe effect \cite{SAWO67}) on a Minkowski background, with the conventional PTA treatment restricting to tensor perturbations \cite{ESWA75,SAZHI78,DETWE79} (see, e.g., \cite{MAGGI07,BUTAET19} for reviews). A fully nonlinear formulation is nevertheless required to consistently describe the observable beyond this regime. Nonlinear gravitational dynamics generate higher-order perturbations and couple different metric sectors, with scalar-induced gravitational waves providing a prominent example \cite{ANCLWA07,BASTET07}. Moreover, beyond linear order the identification of gravitational waves with transverse-traceless metric perturbations becomes gauge dependent, making an operational definition in terms of an observable particularly important \cite{HWJENO17,DOPIWA26}.

Many authors have investigated nonlinear gravitational effects in pulsar timing \cite{NANO23new,CAHEET23,INKOTE24,FRIOET23}, although in most cases the timing response itself is still evaluated at linear order. At second order, the observable receives additional contributions from perturbations of the photon propagation, the emission and observation events, the observed direction, and the pulsar and observer motion, together with the higher-order metric perturbations. 
Explicit treatments of the timing modulation beyond linear order remain limited, with the notable exception of the recent work \cite{DOPIWA26}, which provides the most systematic formulation to date. While substantially advancing the nonlinear description, that analysis focuses on the quadrupolar gravitational-wave contribution to the second-order observable, rather than deriving the complete second-order timing modulation. A fully general second-order treatment therefore remains to be established.

The theoretical description of the observables such as the pulsar timing residuals should be gauge invariant, and the verification of its gauge invariance provides a powerful way to test the sanity of the theoretical description \cite{YOFIZA09,YODU17,MIYO20,MAYO22}. A subtle aspect that is often neglected in the literature is the distinction between two independent coordinate systems: the spacetime coordinates used to describe propagation from the source to the observer, and the observer coordinates used to describe observables such as the angular position of the radio pulsar and the pulsar timing residuals. In particular, the metric tensor in the spacetime relevant for the PTA observations is perturbed around the Minkowski metric ($g_{\mu\nu}=\eta_{\mu\nu}+h_{\mu\nu}$), while the metric tensor in the observer rest frame is exactly the Minkowski metric. Hence at the background level the distinction between two coordinates vanishes (which is the reason behind the negligence). However, these two coordinates are independent, and  the latter is a coordinate system of the tangent space at the observer spacetime position, not of the full spacetime (see \cite{MIYO20} for details). Therefore, the theoretical description of the observables should be independent of spacetime coordinates and gauge-invariant, but of course it transforms accordingly with respect to the change of the observer coordinate. In this respect, still lacking in the literature is an explicit verification of the gauge invariance of the theoretical description for the PTA observables beyond the linear order in perturbations.

In this work, we first formulate the pulsar timing modulation fully nonlinearly in terms of the proper-time intervals between successive emissions and receptions of radio pulses. We then derive its complete second-order expression around a Minkowski background in the limit of an infinitesimal emission interval. We retain the full scalar, vector, and tensor content of the metric perturbations, perform the calculation without fixing a gauge, and consistently account for perturbations in the photon propagation, the emission and reception events, and the motion of the pulsar and observer.
We subject the final result to two nontrivial consistency tests. Although the individual perturbative quantities entering the calculation depend on the coordinate choice, the timing modulation is defined entirely in terms of proper-time intervals and must therefore be a coordinate-independent observable. We explicitly demonstrate this property for the complete second-order expression. As a further test, we make use of our recent nonperturbative result showing that, in the infinitesimal emission-interval limit, the pulsar timing modulation is exactly equivalent to the observed redshift defined along a single null geodesic \cite{MAGIYOO26}. For this comparison, we specialize the fully relativistic second-order observed-redshift expression derived in cosmological perturbation theory in Ref.~\cite{MAYO22} to the present setup. The two results agree exactly, providing a stringent check that all second-order contributions have been consistently included.

The remainder of this paper is organized as follows. In Section~\ref{sec:nonlin}, we define the geometric setup for PTA observations, introduce the observed pulsar timing modulations as the primary observable, and formulate the fully nonlinear framework for its computation at arbitrary orders in perturbation theory. Section~\ref{sec:first} reviews the first-order computation of the observed pulsar timing modulations, while Section~\ref{sec:second} details the derivation at second order. In Section~\ref{sec:gauge}, we explicitly demonstrate that the resulting second-order expression is independent of the coordinate choice by proving that it is a scalar under spacetime diffeomorphisms. Finally, in Section~\ref{sec:disc} we summarize and discuss our findings. Technical details and supporting derivations are collected in the appendices. In particular, Appendix~\ref{app:z} proves that the explicit second-order expression for the pulsar timing modulation is exactly equivalent to the corresponding second-order frequency redshift defined along a single null geodesic.

\section{Fully Nonlinear Description of the Observed Pulsar Timing Residuals}
\label{sec:nonlin}
We develop a fully nonlinear description for the observed pulsar timing residuals, tracing the light propagation of two consecutive emissions of radio signals from a pulsar. Two radio pulses are emitted along the pulsar worldline, and they are received by the observer along the observer worldline. Two emissions and receptions of the radio pulses are related by two null geodesics. The nonlinear description in this section provides the theoretical foundation for the perturbative calculations in the following sections. The second-order theoretical calculations presented in Sec.~\ref{sec:second} constitute the main focus of this work.

\subsection{Geometric setup for PTA observations}
\label{sec:setup}

We consider the light propagation of electromagnetic radio pulses from a  pulsar in our Galaxy to an observer, who represents radio observations in a pulsar timing array. In the following, we describe this physical configuration within perturbation theory around a Minkowski background. This approximation is justified because the relevant light propagation occurs on scales much smaller than the Hubble radius and within a gravitationally bound system. The spacetime metric~$g_{\mu\nu}$ can therefore be written as
\begin{equation}\label{metric}
    ds^2=-(1+2\A)dt^2-2\BB_i \,dx^i dt + (\delta_{ij}+2^{}\CC_{ij})dx^i dx^j \,,
\end{equation}
where calligraphic letters denote spacetime fields describing deviations from the Minkowski metric, and Cartesian coordinates have been assumed so that the spatial background metric is given by the Kronecker delta~$\delta_{ij}$. At this stage, the spacetime fields~$\A$,~ $\BB_i$, and~$\CC_{ij}$ are not required to be small, and therefore no perturbative expansion is performed. Furthermore, the line element does not correspond to any particular gauge choice and retains the full ten independent components of the metric. Throughout this work, an overdot denotes differentiation with respect to the coordinate time~$t$, while a comma denotes differentiation with respect to the spatial coordinates~$x^i$.

\begin{figure}[H]
    \centering
\includegraphics[width=0.85\textwidth]{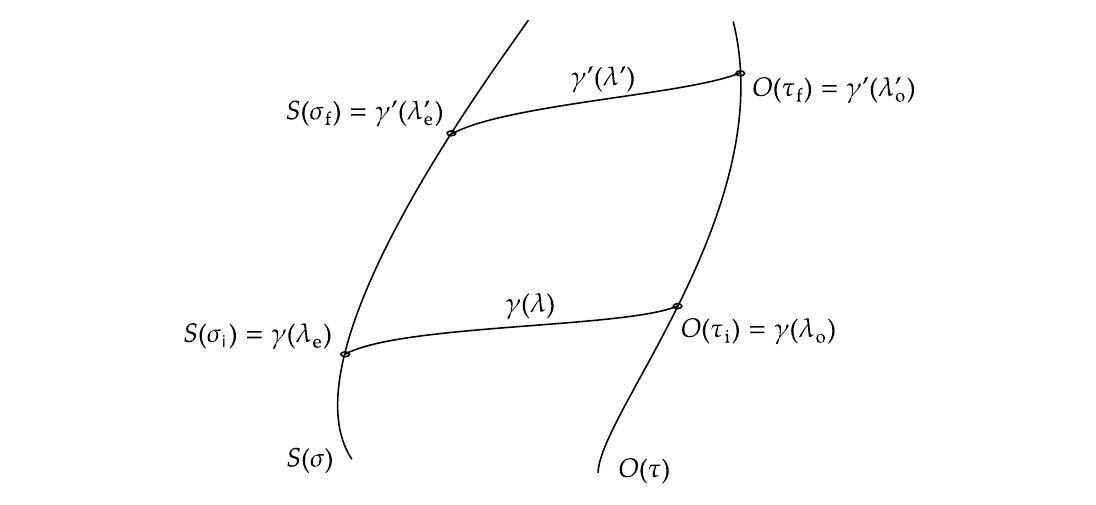}
    \caption{Illustration of the geometric setup: ${O}(\tau)$ is the observer worldline parametrized by its proper time~$\tau$, while ${S}(\sigma)$ is the pulsar worldline parametrized by~$\sigma$. The propagation of two radio pulses is described by two null geodesics~$\gamma(\lam)$ and $\gamma'(\lam')$ with the corresponding affine parameters~$\lam$ and~$\lam'$.
}
    \label{fig:myfig}
\end{figure}

Figure~\ref{fig:myfig} illustrates the geometric construction for the observations in a pulsar timing array. The observer follows a time-like worldline~$O(\tau)$, parameterized by the observer proper time~$\tau$, while the pulsar follows a time-like worldline~$S(\sigma)$, parameterized by the pulsar proper time~$\sigma$. The initial radio pulse is emitted by the pulsar at proper time~$\sigma_{\rm i}$ and received by the observer at proper time~$\tau_{\rm i}$. The trajectory of this initial radio pulse is described by a null geodesic~$\ga(\lam)$, parameterized by an affine parameter~$\lambda$. The second (final) radio pulse is emitted at proper time~$\sigma_{\rm f}$, after one rotation period~$\mathbb{T}_\emi$ in the pulsar rest frame, and it propagates along a second null geodesic~$\ga'(\lam')$, parameterized by the affine parameter~$\lambda'$. This second radio pulse is received by the observer at proper time~$\tau_{\rm f}$, defining the observed rotation period $\mathbb{T}_\obs\equiv\tau_{\rm f}-\tau_{\rm i}$. The emission and observation events are marked by the intersections of the null geodesics and the worldlines.

To simplify the notation, we use the same symbol~$x^\mu$ to denote the coordinates of spacetime points, which include the observer worldline~$x^\mu(\tau)$, the pulsar worldline~$x^\mu(\sigma)$, and the two null geodesics $x^\mu(\lam)$, $x^\mu(\lam')$. Their parameter dependence unambiguously identifies the corresponding trajectory, and the conditions for the geometric intersections impose the following equalities:
\begin{equation}
    x^{\mu }( \sigma_{\mathrm i}) = x^{\mu }(\lambda_\emi)\,,
    \qquad\qquad\qquad
    x^{\mu }( \tau_{\mathrm i}) = x^{\mu }(\lambda_\obs)\,,
\end{equation}
for the first radio propagation and
\begin{equation}
    x^{\mu }( \sigma_{\mathrm f}) = x^{\mu }(\lambda'_\emi) \,,
    \qquad\qquad\qquad
    x^{\mu }( \tau_{\mathrm f}) = x^{\mu }(\lambda'_\obs)\,,
\end{equation}
for the second radio propagation. The subscripts~$\mathrm{i}$ and~$\mathrm{f}$ are reserved for the timelike worldlines of the source and observer, while~$\emi$ and~$\obs$ denote the emission and observation endpoints of the null geodesics.

We now introduce the tangent vectors to the observer and pulsar worldlines defined in the geometric construction above. These are the corresponding four-velocities,
\begin{equation}\label{velNL}
    \frac{dx^\mu}{d\si}\equiv u^\mu(x_\si)\,,\qquad\qquad\qquad
    \frac{dx^\mu}{d\sigma}\equiv u^\mu(x_\sigma)\,,\qquad\qquad\qquad 
u_\mu u^\mu=-1\,.
\end{equation}
No assumption of geodesic motion is made for either the observer or the pulsar. Although the two four-velocities are different, we denote them by the same symbol~$u^\mu$, since the path on which the vector is evaluated removes any potential ambiguity. For the light propagation, however, we keep the two tangent vectors explicitly distinguished in their notation:
\begin{equation}\label{kNL}
    \frac{dx^\mu}{d\lambda}\equiv k^\mu(x_\lambda)\,,
    \qquad\qquad\qquad
    \frac{dx^\mu}{d\lambda'}\equiv k'^\mu(x_{\lambda'})\,,
    \qquad\qquad\qquad
    k^{(\prime)}_\mu k^{(\prime)\mu}=0 \,.
\end{equation}
Here~$k^\mu$ and~$k'^\mu$ are the photon wave vectors associated with the first and second radio pulses, respectively. Since both  pulses are described by null geodesics, their tangent vectors satisfy the geodesic equation:
\begin{equation}\label{geox}
    k^\nu\nabla_\nu k^\mu=0\,,
    \qquad\qquad\qquad
    k'^\nu\nabla_\nu k'^\mu=0 \,.
\end{equation}

Having defined our notation convention for the spacetime coordinates, we want to relate these spacetime coordinates to the physical observables in a pulsar timing array. First, integrating the four-velocity in Eq.~\eqref{velNL} along the worldline yields the coordinate separations~$\Delta x^\mu$ between two emissions
\begin{equation}\label{DxNL}
\Delta x^\mu_\emi\equiv x^\mu(\sigma_{\rm f})-x^\mu(\sigma_{\rm i})=
\int_{\sigma_{\rm i}}^{\sigma_{\rm f}}d\sigma\,u^\mu(x_\sigma)
=x^\mu(\lam'_\emi)-x^\mu(\lam_\emi)\,,
\end{equation}
and~$\Delta x^\mu_\obs$ is defined for the coordinate separation between two observations. The coordinate separations defined above are gauge-dependent and therefore do not by themselves represent physical observables. The relevant observables in pulsar timing array observations are instead the times of arrival (ToA) of radio pulses and the time intervals (or observed rotation period~$\mathbb{T}_\obs$) of two consecutive radio pulses measured in the rest frame of the observer. By construction, these are given by the proper-time intervals separating the two emission and observation events:
\begin{equation}\label{properdiff}
    \mathbb{T}_\emi\equiv\sigma_{\mathrm f} -\sigma_{\mathrm i} \,,
    \qquad\qquad\qquad
    \mathbb{T}_\obs\equiv\tau_{\mathrm f} -\tau_{\mathrm i} \,.
\end{equation}
While the emitted proper-time interval~$\mathbb{T}_\emi$ (or pulsar rotation period) is fixed by the intrinsic properties of the pulsar, the observed interval~$\mathbb{T}_\obs$ between successive radio pulses is affected by fluctuations accumulated during the propagation of the signal from the pulsar to the observer, such as those induced by gravitational waves. The resulting timing modulation\footnote{The primary observables in PTA observations are the times of arrival of radio pulses. Given the proper-time interval between successive emissions, the corresponding observed proper-time interval is modulated by spacetime fluctuations, including gravitational waves. This timing modulation~$z$ in Eq.~\eqref{zdef} is sometimes referred to as the observed redshift in PTA observations, although this terminology is somewhat imprecise. The term ``redshift'' conventionally refers to the change in photon wavelength or frequency between emission and observation, whereas PTA timing measurements are based on the intervals between successive pulses. Strictly speaking, $\mathbb{T}_\obs>\mathbb{T}_\emi$ corresponds to a redshift, while $\mathbb{T}_\obs<\mathbb{T}_\emi$ corresponds to a blueshift. Nevertheless, as shown in Ref.~\cite{MAGIYOO26}, the theoretical descriptions of the pulsar timing modulation and the observed redshift become exactly equivalent in the limit of a vanishing emission interval, $\mathbb{T}_\emi\rightarrow0$, providing a justification for the misnomer.} (or ``observed redshift''~$z$) is then defined as the fractional change in the observed pulse period:
\begin{equation}\label{zdef}
    1+z\equiv\frac{\mathbb{T}_{\obs}}{\mathbb{T}_{\emi}}\,,
\end{equation}
and the observed pulsar timing residual is the accumulation of~$z$ in time from a reference time~$\tau_{\rm ref}$: \begin{equation}\label{PTAresidual}
r(\tau)\equiv\int_{\tau_{\rm ref}}^\tau~d\tau'~z(\tau')~, 
\end{equation}
where the observer proper-time~$\tau$ is the time coordinate the observer uses in the rest frame. While the pulsar rotation period~$\mathbb{T}_\emi$ is a priori unknown, it can be modeled in determining the observed pulsar timing residuals~$r(\tau)$ \cite{SAZHI78,DETWE79,HEDO83,NANO23new,NANO23gw,EPTA23C,EPTA23d,BAHE86,CAFI18,MAGGI18}, once the times of arrival data are collected over a sufficiently long period. For our purposes here, we simply assume that $\mathbb{T}_\emi$ is known, just as the absolute luminosity is assumed to be known in computing the luminosity distance fluctuations \cite{SASAK87,BODUGA06,HUGR06,YOSC16} in supernova observations.

Finally, to relate the coordinate separations to the physical rotation period~$\mathbb{T}_\emi$ and the observed time interval~$\mathbb{T}_\obs$ measured in the pulsar and the observer rest frames, we project the spacetime separations $\Delta x^\mu$ along their local rest frame, defined by the corresponding four-velocities~$u^\mu$ in Eq.~\eqref{velNL}. Expanding the integral for~$\Delta x^\mu$ in Eq.~\eqref{DxNL} in powers of~$\mathbb{T}_\emi$, we obtain
\begin{equation}\label{period}
   ( g_{\mu\nu}\De x^\mu u^\nu)_\emi=-\mathbb{T}_\emi+\mathcal{O}(\mathbb{T}_\emi^2)\,,
\end{equation}
with an analogous relation at observation. Hence, the observed timing modulation~$z$ is recast as
\begin{equation}\label{zcoord}
    1+z={\mathbb{T}_\obs\over\mathbb{T}_\emi}=
\frac{(g_{\mu\nu}\De x^\mu u^\nu)_{\obs}}{(g_{\mu\nu}\De x^\mu u^\nu)_{\emi}}
+\mathcal{O}(\mathbb{T}_\emi)\,.
\end{equation}
In the limit of vanishing rotation periods, $\mathbb{T}_\emi\to0$, only the terms proportional to the pulse periods in Eq.~\eqref{period} and its observation analogue contribute to Eq.~\eqref{zcoord}. Terms involving higher powers of the pulse periods correspond to finite-period corrections and are neglected throughout this work, both for the pulsar timing modulation~$z$ and for the pulsar timing residual~$r(\tau)$ \cite{MAGIYOO26}. This approximation of the vanishing rotation period is a physical assumption of the PTA observations, distinct from the perturbative expansion in the metric and velocity perturbations introduced below. Our main objective is to take this approximation and to evaluate Eq.~\eqref{zcoord} to second order in the metric and velocity perturbations.

\subsection{Propagation of radio pulses along null geodesics}

The expression derived in Eq.~\eqref{zcoord} relates the pulsar timing modulation~$z$  to the infinitesimal coordinate separations~$\Delta x^\mu_\emi,\Delta x^\mu_\obs$ between two radio pulses at the emission and observation events, respectively. They cannot be determined from the observer and pulsar worldlines alone. Additional information is provided by the fact that the emission and observation events of each radio pulse are connected by a null geodesic. We now incorporate this constraint by solving the null geodesic equations associated with the two light rays.

Here we solve the geodesic equation to obtain the spacetime coordinate~$x^\mu_\lam$ along the photon path (see \cite{YOO14a,YOGRET18,MAYO22} for the details and our notation convention). We begin by introducing the nonlinear parameterization of the photon wave vectors defined in Eq.~\eqref{kNL},
\begin{equation}\label{ksplit}
    k^\mu\equiv\left(1+\delta \nu ,\, -\bar n^i-\delta n^i\right)^\mu \,,
\qquad\qquad\qquad \delta_{ij}\bar n^i\bar n^j=1~,
\end{equation}
where the Latin indices $i,j,k,\cdots$ represent spatial coordinates. The dimensionless quantities~$\delta\nu$ and~$\delta n^i$ characterize deviations from the photon propagation in the background Minkowski spacetime. Specifically,~$\delta\nu$ encodes fluctuations in the angular frequency, while~$\delta n^i$ encodes the deviation of the propagation direction from the background direction~$\bar n^i$ (opposite to the propagation direction). Note that the affine parameter has been rescaled with the observed frequency such that the background photon angular frequency is normalized to unity, so that any difference in a real universe is captured by the perturbations~$\delta\nu$ and~$\delta n^i$. For concreteness, the construction  in this subsection will be carried out for the first radio pulse described by~$x^\mu_\lam$. However, the derivation applies identically to the second radio pulse propagating along the null geodesic~$x_{\lambda'}$. Since both radio pulses propagate on the same Minkowski background, they share the same background propagation direction~$\bar n^i$. However, the wave-vector fluctuations~$\delta\nu$ and~$\delta n^i$, as well as the real observed direction~$\bm{\hat n}$, are different for the two radio pulses.

We first integrate the geodesic equations~\eqref{geox} to obtain formal expressions for the components of the photon wave vector at an arbitrary point along the light trajectory:
\begin{equation}\label{geoNL}
    \delta\nu(x_\lambda)=\delta\nu_\obs-\int_{\lambda_\obs}^{\lambda}d\hat\lambda~
    \Gamma^0_{\mu\nu}k^\mu k^\nu \,,
    \qquad\qquad\qquad
    \delta n^i(x_\lambda)=\delta n^i_\obs+\int_{\lambda_\obs}^{\lambda}d\hat\lambda~
    \Gamma^i_{\mu\nu}k^\mu k^\nu \,,
\end{equation}
where~$\Gamma^\mu_{\nu\rho}$ are the Christoffel symbols constructed from the metric~$g_{\mu\nu}$ in Eq.~\eqref{metric}. The integration constants are fixed by boundary conditions imposed at the observer position to match the physical wave-vector in the observer rest frame. These are specified in terms of the observer tetrad~$e^\mu_A$ through the decomposition $k^\mu = e^\mu_A k^A$ \cite{YOGRET18,MIYO20},
\begin{equation}\label{boundary}
    \delta\nu_\obs=e^0_A(1,-\bm {\hat n})^A-1 \,,
    \qquad\qquad\qquad
    \delta n^i_\obs=-e^i_A(1,-\bm {\hat n})^A-\bar n^i \,.
\end{equation}
Here~$\bm {\hat n}$ denotes the observed line-of-sight direction and should not be confused with~$\bm {\bar n}$. The former is observable, whereas the latter is a background quantity introduced for the perturbative expansion. Here and in the following, capital Latin indices denote tetrad components; additional details are given in Appendix~\ref{app:k}.

A further integration of the photon wave vector determines the coordinate difference between the emission and observation events in terms of the geodesic solution in Eqs.~\eqref{geoNL} and \eqref{boundary}:
\begin{equation}\label{coordsNL}
    t_\emi -t_\obs=-d+\int_{\lambda_\obs}^{\lambda_\emi} d\lambda ~ \delta \nu ( x_{\lambda }) \,,
    \qquad\qquad\qquad
    x_\emi^{i } -x_\obs^{i }=\bar n^{i }d-\int_{\lambda_\obs}^{\lambda_\emi} d\lambda~\delta n^{i}( x_{\lambda }) \,,
\end{equation}
where we have defined the affine distance to the pulsar as
\begin{equation}
    d \equiv \lambda_\obs -\lambda_\emi \,.
\end{equation}
Combining the last two equations allows the affine distance to be eliminated, yielding a relation between the temporal coordinate and the radial spatial coordinate
\begin{equation}\label{main}
    t_\emi-t_\obs=x_{\parallel\obs}-x_{\parallel\emi}
+\int_{\lambda_\obs}^{\lambda_\emi} d\lambda ~\Big[\delta \nu ( x_{\lambda })-\delta n_\parallel( x_{\lambda })\Big] \,.
\end{equation}
Here and in the following, the subscript~$\parallel$ denotes the projection of a spatial index along the background propagation direction~$\bar n^i$. Since~$\bar n^i$ is a fixed background vector, it remains constant along the photon trajectory, allowing the projection to be performed under the integral sign.

A useful form of the integrand is obtained by exploiting the null condition of the photon wave vector,~$k_\mu k^\mu=0$, which yields the fully nonlinear relation \cite{YOZA14,MAYO22} as
\begin{align}\label{null}
    \delta\nu -\delta n_\parallel&=-\A+\BB_\parallel+\CC_{\parallel\parallel}+\delta\nu\BB_\parallel+2\CC_{\parallel i}\delta n ^i-2\A\delta\nu+\BB_i\delta n^i-\frac12\delta\nu^2+\frac12 \delta n_i \delta n^i
    \nonumber\\
    &\quad
    -\A \delta\nu^2+\delta\nu \BB_i\delta n^i+\CC_{ij}\delta n^i \delta n^j \,.
\end{align}
To maintain a compact notation in the remainder of the text, we define
\begin{equation}\label{PHI}
    \Phi\equiv\delta \nu-\de n_\parallel\,.
\end{equation}

\subsection{Observed pulsar timing residuals}

Having integrated the null geodesic equations, we now return to the geometric construction introduced at the beginning of this section and compute the infinitesimal coordinate separations that determine the observed pulsar timing residuals. Starting from Eq.~\eqref{main}, we derive the corresponding relation for the second radio pulse and then take the difference in the expressions for two radio pulses. This yields
\begin{equation}\label{Dttemp}
    \Delta t_\emi-\Delta t_\obs=\Delta x_{\parallel\obs} -\Delta x_{\parallel\emi}+\int _{\lambda' _\obs}^{\lambda' _\emi} d\lambda' ~ \Phi ( x_{\lambda'} )-\int _{\lambda _\obs}^{\lambda _\emi} d\lambda ~ \Phi ( x_{\lambda }) \,.
\end{equation}
A subtlety exists in the fact that the affine parameters~$\lambda$ and~$\lambda'$ parameterize two distinct null geodesics. In other words, two line-of-sight integrations over~$d\lam$ and~$d\lam'$ are performed along two different geodesic paths. Furthermore, the corresponding affine distances between emission and observation need not coincide in the physical spacetime ($d\neq d'$). The two affine distances coincide in the Minkowski background, where the temporal coordinate separations for emission and observation events are equal to each other $\Delta t_\emi=\Delta t_\obs\equiv\widebar{\Delta t}$, corresponding to the physical rotation period of a pulsar and the observed time interval ($\widebar{\Delta t}=\mathbb{T}_\emi=\mathbb{T}_\obs$). Note that the observer and the pulsar stay in the same spatial coordinates in the Minkowski background ($\Delta x^i_\emi=\Delta x^i_\obs=0$).

Since affine parameters are defined only up to affine transformations ($\lambda\rightarrow c_1\lambda+c_2$), we use the remaining additive freedom in~$c_2$ to choose a common origin for the two geodesics, $\lambda_\obs=0=\lambda'_\obs$. We recall that the multiplicative freedom in~$c_1$ has already been fixed in the previous subsection by the frequency normalization at the observation positions~$k^A_{\obs}=(1,-\bm{\hat n})^A$ and~$k'^A_{\obs'}=(1,-\bm{\hat n'})^A$; hence no more residual affine transformation is possible. If the emission point of the first radio pulse is located at~$\lambda_{\emi}$, then the emission point of the second radio pulse is located at $\lambda'_\emi=\lambda_\emi+\delta\lambda_\emi$, where we defined
\begin{equation}
    \delta\lambda_\emi\equiv d-d' =\lam'_\emi-\lam_\emi\,.
\end{equation}
The difference between the affine endpoints can be obtained directly from Eq.~\eqref{coordsNL}. Evaluating the relation for both pulses and subtracting the resulting expressions, we find
\begin{equation}\label{delam}
        \delta\lambda_\emi=\Delta x_{\parallel\obs} -\Delta x_{\parallel\emi}-\int _{0}^{\lambda_\emi+\delta\lambda_\emi} d\lambda ' ~\delta n_\parallel( x_{\lambda '} )+\int _{0}^{\lambda_\emi} d\lambda ~ \delta n_\parallel ( x_{\lambda })\,,
\end{equation}
which can be solved order by order in perturbation theory.

Equation \eqref{delam} still depends on the radial coordinate separations~$\Delta x_\parallel$ between successive emission and observation events. We now express these intervals in terms of the observer and pulsar four-velocities. To this end, we introduce the following nonlinear decomposition of the timelike vector field at a generic spacetime point:
\begin{equation}\label{splitu}
    u^\mu\equiv\left(1+\delta u ,\, \UU^i\right)^\mu \,.
\end{equation}
Here~$\UU^i$ denotes the peculiar velocity, which vanishes in the Minkowski background. The temporal fluctuation~$\delta u$ is not an independent degree of freedom and is determined by the normalization condition~$u_\mu u^\mu=-1$, which yields the fully nonlinear relation:
\begin{equation}\label{normu}
    0 = \delta u + \mathcal{A}(1 + 2\delta u) + \frac{1}{2}\delta u^2(1 + 2\mathcal{A}) + \mathcal{B}_i \UU^i(1 + \delta u) - \frac{1}{2}\UU_i \UU^i-\UU^i \UU^j\mathcal{C}_{ij}\,.
\end{equation}
Using Eq.~\eqref{DxNL} and retaining only terms linear in the proper-time intervals~$\mathbb{T}_\emi,\mathbb{T}_\obs$ consistent with the assumption of a vanishing rotation period, we obtain
\begin{equation}\label{Dxpara}
    \Delta x_{\parallel\obs} -\Delta x_{\parallel\emi}=\mathbb{T}_\obs\,\UU_{\parallel\obs}-\mathbb{T}_\emi \,\UU_{\parallel\emi}
    \,,
\end{equation}
where we defined $\UU_{\parallel\obs}\equiv\UU_\parallel(x_{\tau_{\rm i}})$ and $\UU_{\parallel\emi}\equiv\UU_\parallel(x_{\sigma_{\rm i}})$ corresponding to the radial peculiar velocities evaluated at the observer and emission events of the first radio pulse, respectively.

Combining Eqs.~\eqref{Dttemp}, \eqref{delam}, and \eqref{Dxpara}, together with our choice of affine parameterization, yields an explicit expression for the temporal coordinate separations between the two radio pulses. Expanding the resulting integrals in the limit $|\delta\lambda_\emi|\ll|\lambda_\emi|$, we obtain the fully nonlinear relation
\begin{equation}\label{Dt}
    \Delta t_\emi-\Delta t_\obs=\Delta x_{\parallel\obs} -\Delta x_{\parallel\emi}+\delta\lam_\emi\Phi(x_{\lam'=\lam_\emi})
    +\int _{0}^{\lambda_\emi} d\lambda' ~ \Phi ( x_{\lambda'} )-\int _{0}^{\lambda_\emi} d\lambda ~  \Phi ( x_{\lambda }) \,.
\end{equation}
In deriving the expression above, we have retained only leading-order terms in~$\delta\lambda_{\emi}$, which is proportional to~$\mathbb{T}_\emi$ in the lowest order. This approximation reflects the geometric construction presented in Sec.~\ref{sec:setup}: the two radio pulses are regarded as neighboring light rays connecting infinitesimally separated emission and observation events. Equivalently, the coordinate separations~$\Delta x^\mu$ and the affine shift~$\delta\lambda_{\emi}$ are treated as infinitesimal quantities, so that higher-order corrections in these separations are consistently neglected.

It is important to emphasize that this approximation is unrelated to perturbation theory. No expansion has been performed in the metric or matter fields, and the dependence on the spacetime geometry remains fully nonlinear. The results obtained in this subsection therefore provide the exact infinitesimal coordinate separations $\Delta x^\mu$ required to compute the observed time interval~$\mathbb{T}_\obs$ in Eq.~\eqref{properdiff} via Eq.~\eqref{period} and to relate it to the pulsar rotation period~$\mathbb{T}_\emi$ for the observed pulsar timing residuals in Eq.~\eqref{PTAresidual} via Eq.~\eqref{zcoord}. Perturbative expansions will be performed starting in the following sections.

\section{Linear-Order Perturbative Calculations}
\label{sec:first}

Before turning to the second-order computation, we first derive the linear-order result by using the nonlinear formalism developed in the previous section. Although the theoretical description of the PTA observations is well established at the linear order in perturbations, the existing derivations are typically restricted to tensor metric perturbations in the transverse-traceless gauge \cite{ESWA75,SAZHI78,DETWE79}. Here, by contrast, we do not fix a gauge and retain the full set of ten independent metric components, including scalar, vector, and tensor modes. This provides a crucial way of checking the validity of the theoretical description \cite{YOFIZA09,YODU17,MIYO20,MAYO22}.

To evaluate the observed timing modulation~$z$ in Eq.~\eqref{zcoord}, we first determine the infinitesimal coordinate separations at first order. The radial separation appearing in Eq.~\eqref{Dxpara} is proportional to the peculiar velocity, which is itself a first-order quantity. It is therefore sufficient to evaluate the periods at background order. In the Minkowski background, any spatial motion is absent and the only nontrivial evolution occurs along the time direction. The background proper-time intervals measured by the pulsar and the observer rest frames are then identical, and Eq.~\eqref{period} reduces to
\begin{equation}
    \mathbb T_\obs=\widebar{\Delta t}+\mathcal{O}(1) \,,
    \qquad\qquad\qquad
    \mathbb T_\emi=\widebar{\Delta t}+\mathcal{O}(1) \,,
\end{equation}
where~$\mathcal{O}(1)$ denotes quantities of first and higher order in the metric and matter perturbations. We thus obtain the first-order expression for the radial separation
\begin{equation}\label{Dxpara1}
    {\Delta x}_{\parallel\obs} -{\Delta x}_{\parallel\emi}=-\widebar{\Delta t}~~
    \UU_{\parallel}\big\rvert^\emi_\obs+\mathcal{O}(2)
    \,.
\end{equation}
Here and in the following, we use the vertical-bar notation to denote the difference between a quantity evaluated at the emission and observation events of the first radio pulse.

To determine the temporal separation at first order, we expand the nonlinear expression in Eq.~\eqref{Dt} and retain only terms linear in the perturbations. At this order, the shift in the affine parameter~$\delta\lambda_\emi$ does not contribute. The integrands in Eq.~\eqref{Dt} are already perturbative quantities. It is therefore sufficient to evaluate the line integrals along the background null geodesics. These are obtained by specializing Eq.~\eqref{coordsNL} to the Minkowski background, yielding
\begin{equation}\label{path0}
    \bar x^\mu(\lambda)=\left( \lambda+\bar t_\obs ,\, -\bar n^i\lambda \right)^\mu \,,
    \qquad\qquad\qquad
    \bar x^\mu(\lambda')=\left( \lambda'+\bar t'_\obs ,\, -\bar n^i\lambda' \right)^\mu \,,
\end{equation}
where we have chosen the background spatial position of the observer to coincide with the origin of the coordinate system.

The affine parameters~$\lambda$ and~$\lambda'$ merely label the two background trajectories and act as dummy integration variables in Eq.~\eqref{Dt}. They may therefore be identified without loss of generality. The two background paths nevertheless differ by a constant temporal shift~$\widebar{\Delta t}$, since the second radio pulse is received a background period~$\widebar{\Delta t}$ after the first. Consequently, the integrand evaluated along the second trajectory can be expanded around that of the first, leading to
\begin{equation}\label{Phidot}
    \Phi\left( \lambda+\bar t'_\obs ,\, -\bar n^i\lambda \right)=\Phi\left( \lambda+\bar t_\obs ,\, -\bar n^i\lambda \right)+\widebar{\Delta t}~~\dot{\Phi}\left( \lambda+\bar t_\obs ,\, -\bar n^i\lambda \right)
    \,.
\end{equation}
Terms of order~$\mathcal{O}(\widebar{\Delta t}{\,}^2)$ have been neglected, since we work in the limit of infinitesimal separations ($\mathbb{T}_\emi\rightarrow0$) and retain only contributions linear in the period~$\widebar{\Delta t}$. Substituting Eqs.~\eqref{Dxpara1} and \eqref{Phidot} into Eq.~\eqref{Dt}, we then obtain an explicit expression for the first-order temporal separation:
\begin{equation}\label{Dt1}
    \Delta t_\obs-\Delta t_\emi=\widebar{\Delta t}~\left[\UU_{\parallel}\big\rvert^\emi_\obs-\int _{0}^{\lambda_\emi} d\lambda ~  \dot\Phi ( \bar x_{\lambda }) +\mathcal{O}(2)\right]\,.
\end{equation}

We now have all the ingredients required to evaluate the observed timing modulation~$z$ in Eq.~\eqref{zcoord} at first order. Expanding Eq.~\eqref{period} to linear order shows that only the temporal coordinate separation and the lapse perturbation contribute at this order:
\begin{equation}\label{lapse}
\mathbb{T}_\emi=-\left(g_{00}~\Delta t~u^0\right)_\emi+\mathcal{O}(2)
~.
\end{equation}
Using the expressions derived above, we therefore find
\begin{equation}\label{z1}
    z=(\UU_\parallel-\A)\big\rvert^{\emi}_{\obs}+\int _{\bar t_\obs}^{\bar t_\emi} d\bar t  ~(\dot\A-\dot\BB_\parallel-\dot\CC_{\parallel\parallel})+\mathcal{O}(2)\,,
\end{equation}
where we have substituted the first-order expression for~$\Phi$ obtained from the nonlinear relation in Eq.~\eqref{null}. We have also replaced the integration over the affine parameter by an integration over the corresponding background coordinate using Eq.~\eqref{path0}, thereby making explicit that the integration is performed along the background null geodesic.  At this order, the subscript~$\parallel$ denotes projection along the observed line of sight. The choice between the line-of-sight directions of the first and second radio pulses is irrelevant at linear order, since the difference between~$\bm{\hat n}$ and~$\bm{\hat n'}$ is already first order and is absorbed into the~$\mathcal{O}(2)$ corrections.

The first term in Eq.~\eqref{z1} corresponds to the Doppler contribution, followed by the Sachs--Wolfe and integrated Sachs--Wolfe terms. We emphasize that this expression is identical to the well-known Sachs-Wolfe formula \cite{SAWO67} in cosmology, i.e., the linear-order expression for the observed redshift obtained from the propagation of a single light ray between the source and the observer. The equivalence is not only for tensor perturbations, but also for scalar and vector perturbations.

\section{Second-Order Perturbative Calculations}
\label{sec:second}
In this section, we turn to the main objective of this work and derive the observed timing modulation~$z$ at second order in relativistic perturbation theory. Although the calculation follows the same conceptual steps as the first-order analysis, its implementation is considerably more involved and requires careful treatment of several subtleties, which we discuss below (see also \cite{MAYO22} for in-depth discussions).

For clarity, we first derive an \emph{implicit} expression for~$z$, in which the infinitesimal coordinate separations are left unevaluated. The resulting formula separates into linear terms, which have the same structure as in the first-order calculation, and quadratic terms involving products of infinitesimal spacetime intervals with metric and matter perturbations.

To obtain the corresponding \emph{explicit} expression, all the quantities must be substituted by their perturbative expansions. Since the first-order infinitesimal separations have already been derived, the only remaining ingredient is the second-order temporal separation. Maintaining this distinction between implicit and explicit expressions not only streamlines the derivation but also facilitates the verification of the diffeomorphism invariance presented in the next section.

\subsection{Implicit expression for the observed pulsar timing modulation~$z$}

In this subsection, we derive the implicit expression for the observed timing modulation~$z$. By implicit, we mean that the perturbative expressions for the infinitesimal coordinate separations are not substituted, but instead left in symbolic form throughout the calculation. To this end, we first evaluate the pulsar rotation period~$\mathbb{T}_\emi$ and its observed time interval~$\mathbb{T}_\obs$. We begin from Eq.~\eqref{period}, written without specifying whether it is evaluated at the observer or the pulsar position, since the same relation applies in both cases. Using also the normalization condition for the four-velocity in Eq.~\eqref{normu}, a direct computation yields
\begin{equation}\label{T}
    \mathbb{T}=\Big[1+\A+ \frac12 (  \UU^2-\mathcal{A}{}^2)\Big] \Delta t- \Delta x^{i} (\UU_i - \mathcal{B}_i) \,.
\end{equation}
Hereafter, any correction terms of $\mathcal{O}(3)$ will be omitted.

Given the expressions for the pulsar rotation period~$\mathbb{T}_\emi$ and its observed time interval~$\mathbb{T}_\obs$, 
the second-order expression for~$z$ follows directly from its definition in Eq.~\eqref{zdef}
\begin{equation}\label{zimp}
    z={\mathbb{T}_\obs\over\mathbb{T}_\emi}-1=
    \Theta-\mathcal{A}\Big\rvert^\emi_\obs(1+\Theta-\A_\emi ) +\frac{1}{2} (\mathcal{A}^2-\UU_i \UU^i)\Big\rvert^\emi_\obs+\frac{1}{\widebar{\Delta t}}\left[\Delta x^{i} (\UU_i - \mathcal{B}_i)\right]\Big\rvert^\emi_\obs \,.
\end{equation}
To simplify the notation, we have introduced the shift~$\Theta$ in the temporal coordinate separations,
\begin{equation}
    1+\Theta\equiv\frac{\Delta t_\obs}{\Delta t_\emi} ~,
\end{equation}
which is defined analogously to the pulsar timing modulation~$z$, but with coordinate-time intervals $\Delta t$ replacing proper-time intervals $\mathbb{T}$. Unlike~$z$, $\Theta$~is therefore not an observable quantity, but merely a useful intermediate variable.

Since~$\Theta$ vanishes at background order, its perturbative expansion up to second order reads
\begin{equation}\label{Dtrel}
\widebar{\Delta t}~ \Theta^{(1)}=\Delta t^{(1)}_\obs-\Delta t^{(1)}_\emi~,
    \qquad\qquad\qquad
\widebar{\Delta t}~ \Theta^{(2)}=\Delta t^{(2)}_\obs-\Delta t^{(2)}_\emi
-\Delta t^{(1)}_\emi~\Theta^{(1)} \,.
\end{equation}
The first-order expression for the temporal shift~$\Theta^{(1)}$ was derived in the previous section (see Eq.~\eqref{Dt1}). It therefore remains to determine the second-order contribution~$\Theta^{(2)}$, which follows from the temporal separations~$\Delta t^{(2)}_\emi-\Delta t^{(2)}_\obs$ through Eq.~\eqref{Dtrel}. Once~$\Theta^{(2)}$ is obtained, its substitution into Eq.~\eqref{zimp} immediately yields the explicit second-order expression for the pulsar timing modulation~$z$.

\subsection{Explicit expression for the observed pulsar timing modulation~$z$}

We now turn to the computation of the second-order observed pulsar timing modulation~$z$. This requires the evaluation of the shift in the temporal separation~$\Theta$ in Eq.~\eqref{Dtrel} at second order, since the first-order contributions entering Eq.~\eqref{T} have already been derived.

We first evaluate the radial separation from Eq.~\eqref{Dxpara}. Since the peculiar velocity is already first order, the second-order contribution is obtained by combining it with the expression for the proper-time interval~$\mathbb{T}$ in Eq.~\eqref{T}. Reading its first-order relation to the temporal separation~$\Delta t$ in Eq.~\eqref{T}, the second-order radial separation is given by
\begin{equation}\label{Dxpara2}
    \Delta x_{\parallel\obs}-\Delta  x_{\parallel\emi}=-\left[\Big(\Delta t+\widebar{\Delta t}~{}^{}\A \Big) \UU_\parallel\right]\bigg\rvert^\emi_\obs \,.
\end{equation}

To complete the derivation of the second-order temporal separation in Eq.~\eqref{Dt}, the remaining ingredients  must also be evaluated. We begin with the integrals of~$\Phi$ along the two light rays, which share the same integration range over the affine parameter. At first order, we evaluated the integral of~$\Phi$ along the two background trajectories and expressed its evaluation along the second pulse in terms of that along the first pulse using Eq.~\eqref{Phidot}. We now follow the same procedure at second order, paying particular attention to the crucial difference that the trajectories must be known to linear order in perturbations. To this end, we first determine the linear-order perturbed photon trajectories at a given affine parameter,
\begin{align}\label{pertpath}
    x^\mu(\lambda)&=\left( \lambda+ t_\obs+\int_{0}^{\lambda} d\hat\lambda ~ \delta \nu\big\rvert_\gamma ~, ~~
-\bar n^i\lambda +x^i_\obs-\int_{0}^{\lambda} d\hat\lambda ~ \delta n^i\big\rvert_\gamma\right)^\mu \,,
    \nonumber\\
    x^\mu(\lambda')&=\left( \lambda'+ t'_\obs+\int_{0}^{\lambda'} d\hat\lambda ~ \delta \nu\big\rvert_{\gamma'} ~,~~
 -\bar n^i\lambda' +x'^i_\obs-\int_{0}^{\lambda'} d\hat\lambda ~ \delta n^i\big\rvert_{\gamma'}\right)^\mu \,,
\end{align}
where~$\big\rvert_{\gamma}$ and~$\big\rvert_{\gamma'}$ denote evaluation along the first and second trajectories of the radio pulses, respectively. In comparing the two trajectories, we identify the dummy integration variables $\lambda'=\lambda$, as in the first-order calculation. The endpoint correction proportional to~$\delta\lambda_e$ has already been isolated in Eq.~\eqref{Dt}.

At fixed~$\lambda$, the second trajectory differs from the first trajectory in three aspects. First, two observation events are separated not only in time, but also in space, while the spatial separation~$\Delta x_\obs^i$ vanishes in the background:
\begin{equation}\label{Dobs}
    t'_\obs=t_\obs+\Delta t_\obs\,,\qquad\qquad x'^i_\obs=x^i_\obs+\Delta x^i_\obs\,.
\end{equation}
Second, the integrals of the photon wave-vector perturbations are evaluated along the corresponding background trajectories. When expressing the evaluation along the second pulse~$\gamma'$ in terms of the first pulse~$\gamma$, the temporal displacement between the two background trajectories generates time derivatives, analogously to the expansion of~$\Phi$ in Eq.~\eqref{Phidot}:
\begin{equation}\label{gamma'gamma}
    \delta \nu\big\rvert_{\gamma'}=\delta \nu\big\rvert_\gamma+\widebar{\Delta t}\,\dot{\delta\nu}\big\rvert_\gamma+\mathcal{O}(2)\,,\qquad\quad\qquad \delta n^i\big\rvert_{\gamma'}=\delta n^i\big\rvert_{\gamma}+\widebar{\Delta t}\,(\dot{\delta n}{}^i+\eta^i)\big\rvert_\gamma+\mathcal{O}(2)\,.
\end{equation}
In the expression for the fluctuation in the photon propagation direction~$\delta n^i$, we find an additional term~$\eta^i$ that is unrelated to the temporal separation between the two trajectories. This term arises from the difference in the observed angles for  two pulses: the first pulse is observed along~$\bm{\hat n}$, while the second is observed along~$\bm{\hat n'}$. The different observed line-of-sight directions lead to different boundary conditions for the corresponding photon trajectories. We parametrize this angular difference as\footnote{We refer the reader to Appendix~\ref{app:k} for details on the notation convention.}
\begin{equation}\label{etadef}
    \widebar{\De t}~\eta^{i}\equiv \delta^i_I(\hat{n}'^I-\hat{n}^I)\,.
\end{equation}
In the limit of infinitesimally separated pulses, the line-of-sight directions~$\bm{\hat n}$ and~$\bm{\hat n'}$ are infinitesimally close, and~$\eta^i$ measures their rate of change. Since both are unit vectors, this variation is orthogonal to the propagation direction. Moreover,~$\eta^i$ vanishes in the background and at first order in perturbations satisfies~$\bar n_i\eta^i=0$. As seen in Eq.~\eqref{gamma'gamma}, this effect contributes only to the fluctuation in the photon propagation direction~$\delta n^i$. Indeed, the observer boundary conditions~$\delta n_\obs^i$ and~$\delta n'^i_\obs$ depend explicitly on the corresponding line-of-sight directions~$\bm{\hat n}$ and~$\bm{\hat n'}$, respectively [see Eq.~\eqref{boundarys1st}]. The analogous dependence of~$\delta\nu$ is already perturbative and therefore contributes only to higher orders.

Taking into account the differences in three aspects identified above when comparing the second pulse with the first, as shown in Eqs.~\eqref{Dobs} and \eqref{gamma'gamma}, we can now express~$\Phi$ in Eq.~\eqref{Dt} evaluated along the second pulse in terms of its evaluation along the first perturbed trajectory~$x_\lam$. This gives the second-order analogue of Eq.~\eqref{Phidot}:
\begin{align}\label{Phiexp}
    \Phi(x_{\lambda'})&=\Phi(x_{\lambda})+\widebar{\Delta t}\,\dot\Phi\big\rvert_{x_\lam}+\widebar{\Delta t}\,\eta^i\left( \BB_i+2\CC_{\parallel i}+\delta n_i \right)\big\rvert_{\bar x_\lam}
    \nonumber\\&\quad
    +(\partial_\mu\Phi)\big\rvert_{\bar x_\lam}\left[ (\Delta t^{(1)}_\obs\,,\,\Delta x^i_\obs)^\mu+\widebar{\Delta t}\int_0^\lam d\hat\lam\,\left(\dot{\delta\nu}\,,\,-\dot{\delta n}{}^i-\eta^i \right)^\mu \right]\,.
\end{align}
Equation~\eqref{Phiexp} organizes the three effects discussed above as follows. The displacement of the observation event in Eq.~\eqref{Dobs} generates the contribution~$(\partial_\mu\Phi)\big\rvert_{\bar x_\lambda}\Delta x_\obs^\mu$. We separate the term associated with the background temporal separation~$\widebar{\Delta t}$ from the first-order separation~$(\Delta t^{\smash{(1)}}_\obs\,,\Delta x_\obs^i)^\mu$, so that the former retains the same form as in the linear-order calculation,~$\widebar{\Delta t}\,\dot\Phi\big\rvert_{x_\lambda}$. At second order, however,~$\dot\Phi$ must be evaluated along the perturbed trajectory of the first pulse; the corresponding expansion around the background trajectory is given in Eq.~\eqref{PhiTWO} below. The relation in Eq.~\eqref{gamma'gamma} generates the integral term in the second line of Eq.~\eqref{Phiexp}. Finally, the change in the line-of-sight direction defined in Eq.~\eqref{etadef} contributes both through Eq.~\eqref{gamma'gamma} and through the term proportional to~$\eta^i$ in the first line, the latter arising from the explicit dependence of~$\Phi$ on~$\delta n^i$:
\begin{equation}\label{Phi2}
    \Phi=-\A+\BB_\parallel+\CC_{\parallel\parallel}+\delta\nu\BB_\parallel+2\CC_{\parallel i}\delta n ^i-2\A\delta\nu
    +\BB_i\delta n^i-\frac12\delta\nu^2+\frac12 \delta n_i \delta n^i
    \,,
\end{equation}
obtained from the second-order expansion of Eq.~\eqref{null}.

As anticipated, we now expand~$\dot\Phi|_{x_\lambda}$ around the background trajectory of the first pulse. Since~$\Phi$ starts at first order, the first-order displacement of the evaluation point generates a second-order contribution. We define this displacement from the background path as
\begin{equation}\label{deltax}
    \delta x^\mu_\lambda
    \equiv x^\mu(\lambda)-\bar x^\mu(\lambda)
    =
    \delta x^\mu_\obs
    +\int_0^\lambda d\hat\lambda~
    (\de\nu\,,-\de n^i)^\mu\,.
\end{equation}
The expansion around the first background pulse then yields
\begin{equation}\label{PhiTWO}
    \dot\Phi\big|_{x_\lambda}
    =
    \left[
    \dot\Phi
    +\delta x^\mu_\lambda(\partial_\mu\dot\Phi)
    \right]_{\bar x_\lambda}\,.
\end{equation}

Using the expansion of the integrals of~$\Phi$ derived above, the temporal coordinate separation in Eq.~\eqref{Dt} now takes the form
\begin{multline}\label{Dt2}
    \Delta t_\obs-\Delta t_\emi=\Delta x_{\parallel\emi} -\Delta x_{\parallel\obs}-\delta\lambda_\emi\Phi_\emi
    -\widebar{\Delta t}\int _{0}^{\lambda_\emi} d\lambda ~\left[\dot\Phi+\delta x^\mu_{\lambda} \,(\partial_\mu \dot\Phi)+\eta^i\left( \BB_i+2\CC_{\parallel i}+\delta n_i \right)\right]\Big\rvert_{\bar x_{\lambda}}
    \\
    -\int_0^{\lam_\emi} d\lam~\bigg[\Big(\Delta t^{(1)}_\obs+\bar{\Delta t}\int_0^\lam d\hat\lam\,\dot{\de\nu}\Big) \dot\Phi\big\rvert_{\bar x_\lam}+\Big(\Delta x^i_\obs-\widebar{\Delta t}\int_0^\lam d\hat\lam\,(\dot{\de n}{}^i+\eta^i)\Big) \Phi_{,i}\big\rvert_{\bar x_\lam}\bigg]\,.
\end{multline}
In deriving this result, we have neglected terms of order~$\mathcal{O}(|\Delta x^\mu|^2)$. Accordingly, at this order we may neglect the distinction between~$\Phi(x_{\lambda'=\lambda_{\rm e}})$ and~$\Phi(\bar{x}_{\lambda_{\rm e}})\equiv\Phi_{\rm e}$.

We now turn to the remaining ingredient entering the temporal fluctuation, namely the shift~$\delta \lam_\emi$ in the affine parameter at the emission events. Since~$\delta\lambda_\emi$ is needed only at first order, it is sufficient to evaluate Eq.~\eqref{delam} to linear order. A more convenient expression, however, is obtained by combining Eq.~\eqref{delam} with the temporal relation in Eq.~\eqref{Dttemp} and using the definition of~$\Theta^{(1)}$ in Eq.~\eqref{Dtrel}. We find
\begin{align}\label{delam1}
    \delta\lambda_\emi&=
-\widebar{\Delta t}\, \Theta^{(1)}
-\widebar{\Delta t}\int _{0}^{\lambda_\emi} d\lambda  ~\dot{\delta \nu}({ \bar x_{\lambda }}) \,.
\end{align}
The last term is obtained by expanding the temporal wave-vector fluctuation along the second pulse around that along the first, using Eq.~\eqref{gamma'gamma}. No contribution proportional to~$\eta^i$ arises, since the temporal fluctuation~$\delta\nu$ has no line-of-sight correction at this order.

Having determined all the contributions to the second-order temporal separation, we can now substitute the radial separation in Eq.~\eqref{Dxpara2} and the affine-parameter shift derived above into Eq.~\eqref{Dt2}, and use the resulting relation together with the definition in Eq.~\eqref{Dtrel} to obtain~$\Theta^{(2)}$. Inserting this explicit result into Eq.~\eqref{zimp} and collecting all terms up to second order, we obtain the observed pulsar timing modulation~$z$:
\begin{align}\label{z2}
    z&= \Big[\UU_\parallel-\A+\UU_\parallel\A+\frac12(\A^2+\UU_i\UU^i)-\BB_i\UU^i\Big]\Big\rvert^{\emi}_{\obs}+(\A_\emi-\Theta^{(1)}) \mathcal{A}\big\rvert^\emi_\obs-\int _{\bar t_\obs}^{\bar t_\emi} d\bar t ~\left[\dot\Phi+\delta x^\mu_{\lambda} \,(\partial_\mu \dot\Phi)\right]
    \nonumber\\&\quad
    +\eta^i\int_{\bar t_\obs}^{\bar t_\emi} d\bar t~\Big[(\bar t-\bar t_\obs)\Phi_{,i}-
    \left( \BB_i+2\CC_{\parallel i}+\delta n_i \right)\Big]-\int_{\bar t_\obs}^{\bar t_\emi} d\bar t~\bigg[(\partial_\mu\Phi)\int_{\bar t_\obs}^{\bar t}d\hat{\bar t}\,(\dot{\de\nu}\,,-\dot{\de n}{}^i)^\mu\bigg]
    \nonumber\\&\quad
    -\UU^i_\obs\int _{\bar t_\obs}^{\bar t_\emi} d\bar t ~\Phi_{,i}+
    \Theta^{(1)}\Big(\Phi_\emi-\UU_{\parallel\obs}-\int _{\bar t_\obs}^{\bar t_\emi} d\bar t ~\dot\Phi\Big)
    +\Phi_\emi\int _{\bar t_\obs}^{\bar t_\emi} d\bar t  ~\dot{\delta \nu}
    \,.
\end{align}
As in the first-order result, we have replaced the affine parameter by the corresponding background time coordinate using Eq.~\eqref{path0}, making manifest that all integrations are performed along the background null geodesic of the first pulse. 

To cast Eq.~\eqref{z2} in a fully explicit form in terms of the metric and velocity perturbations together with the observer boundary conditions, we substitute~$\Theta^{(1)}$ from Eq.~\eqref{Dt1}, the linear-order wave-vector perturbations derived in Appendix~\ref{app:k}, and the second-order form of~$\Phi$ in Eq.~\eqref{Phi2}. For the expression of~$\eta^i$, we introduce the projection operator orthogonal to~$\bar n^i$: $P^i{}_j\equiv\delta^i{}_j-\bar n^i\bar n_j$. Projecting the spatial relation in Eq.~\eqref{coordsNL} transverse to~$\bar n^i$, taking the difference between the two pulses, and using Eq.~\eqref{gamma'gamma} for the difference between the corresponding wave-vector perturbations, we obtain at first order
\begin{equation}
    \Delta x^i_{\perp\emi}-\Delta x^i_{\perp\obs}
    =
    -\widebar{\Delta t}\int_0^{\lambda_\emi}d\lambda\,
    \left(\dot{\delta n}{}^i_\perp+\eta^i\right)\,,
\end{equation}
where the subscript~$\perp$ denotes projection with~$P^i{}_j$. Using the relation between the peculiar velocity and the spatial coordinate separation in Eq.~\eqref{DxNL},
\begin{equation}
    \Delta x^i_{\perp\obs}
    =
    \widebar{\Delta t}~\UU^i_{\perp\obs}\,,
    \qquad\qquad\qquad
    \Delta x^i_{\perp\emi}
    =
    \widebar{\Delta t}~\UU^i_{\perp\emi}\,,
\end{equation}
recalling that~$d=-\lambda_e$, and imposing the first-order transverse condition~$\bar n_i\eta^i=0$, we find
\begin{equation}\label{eta}
    \eta^{i}
    =
    \frac{1}{d}
    \left(
    \UU^i_{\perp}\big\rvert^\emi_\obs
    +\int_0^{\lambda_e}d\lambda\,
    \dot{\delta n}{}^{i}_\perp(\bar x_\lambda)
    \right)\,.
\end{equation}
We emphasize that~$\eta^i$ is not a quantity defined along the photon trajectory and therefore carries no dependence on the affine parameter. Rather, it is a single transverse vector associated with the pair of pulses, encoding the infinitesimal change between their observed line-of-sight directions. Equation~\eqref{eta} expresses this change in terms of the endpoint velocities and the integrated variation of the photon propagation direction along the first trajectory.

Equation~\eqref{z2} is the central result of this work. It gives the observed pulsar timing modulation at second order without fixing a gauge and contains the complete set of local contributions at the emission and observation events, line-of-sight contributions accumulated during photon propagation, and corrections associated with the displacement between the two neighboring photon trajectories and the change in their observed line-of-sight directions. Although several auxiliary quantities were introduced in the derivation, none of them is separately observable; only their combination in Eq.~\eqref{z2} represents the PTA observable, i.e., the observed timing modulations.

In the following section, we perform a geometric consistency check of Eq.~\eqref{z2} by verifying that the second-order result satisfies the scalar transformation law required by a physical observable, providing a necessary condition for the validity of the expression. As a further validation, Appendix~\ref{app:z} shows that the second-order result obtained from the two-pulse construction agrees with the corresponding single-pulse observed redshift derived in \cite{MAYO22}. Although the equivalence is guaranteed by the nonlinear identity established in Ref.~\cite{MAGIYOO26}, recovering it explicitly at second order is highly nontrivial. It requires the exact cancellation and recombination of all auxiliary contributions introduced by the two-pulse construction into the independently derived single-pulse expression. The agreement therefore confirms the internal consistency of both the two-pulse and single-pulse calculations, as well as the completeness of the second-order derivation.

\section{Geometric Consistency Checks}
\label{sec:gauge}
Throughout this work, we have deliberately avoided fixing a gauge. Besides making the theoretical description developed here applicable to arbitrary gauges, this enables a stringent consistency check \cite{YOFIZA09,YODU17,MIYO20,MAYO22} of the complete second-order expression for the observed pulsar timing modulation~$z$ derived in Eq.~\eqref{z2}. Since this quantity represents a physical observable, its value cannot depend on the coordinates used  here to label spacetime points. The consistency check performed in this section therefore consists in verifying that the complete second-order expression transforms as a scalar under infinitesimal spacetime diffeomorphisms.

Our strategy closely parallels the derivation of the second-order expression itself in Sec.~\ref{sec:second}. We first prove that the implicit expression for~$z$ transforms as a scalar, assuming that the coordinate separations transform as coordinate differences. We then verify that the explicit expressions for these separations, derived in the previous section in terms of the metric and matter perturbations, indeed satisfy the required transformation laws. We emphasize that both the derivation and the consistency check are purely geometrical, thus they apply to any metric theory in which electromagnetic pulses propagate on null geodesics of the metric.

\subsection{Diffeomorphisms and gauge transformations}

Although the complete second-order expression derived in Eq.~\eqref{z2} is rather involved, the observed pulsar timing modulation~$z$ was defined geometrically in Eq.~\eqref{zdef} as the ratio of two proper-time intervals. As such, it represents an observable quantity and therefore cannot depend on the coordinate system used to describe the spacetime. More generally, observable quantities must be invariant under arbitrary coordinate transformations \cite{YOFIZA09,YODU17,MIYO20,MAYO22}. Coordinates are only labels assigned to spacetime points by the theorist and therefore cannot affect the outcome of physical measurements such as the pulsar timing residuals measured in PTA observations.

If two coordinate systems are related by an infinitesimal diffeomorphism generated by the vector field~$\xi^\mu$,
\begin{equation}\label{coordtransf}
    \tilde x^\mu(x)=x^\mu+\xi^\mu(x) \,,
\end{equation}
then an observable quantity, such as the observed pulsar timing modulation~$z$, satisfies
\begin{equation}\label{zscalar}
    \tilde z(\tilde x_{\rm p})=z(x_{\rm p}) \,,
\end{equation}
where~$\tilde z$ and~$z$ denote the same quantity expressed in the coordinate systems~$\tilde x^\mu$ and~$x^\mu$, respectively, and are evaluated at the same arbitrary physical spacetime point~$\rm p$, which corresponds to different coordinate values in the two charts, related by Eq.~\eqref{coordtransf}.

A gauge transformation is a more subtle concept than the coordinate transformation described above. In cosmological perturbation theory, a gauge transformation corresponds to a change in the identification map between the background spacetime and the perturbed spacetime describing the real Universe \cite{BARDE80}. Such a change can be represented as a coordinate transformation with the transformed (tilde) and untransformed (un-tilde) quantities compared at the same coordinate value, which therefore corresponds to two different physical points. As a consequence, invariance under spacetime diffeomorphisms of the full quantity and gauge invariance of its perturbations are, in general, distinct notions. At linear order, this distinction disappears for perturbations of quantities with a constant background value. Beyond linear order, however, the gauge transformation of a given perturbative order also involves perturbations of lower orders, and the two notions no longer coincide. This correspondence can nevertheless be restored by defining perturbations with respect to a coordinate-independent reference point in the background spacetime. In that case, quantities that transform as scalars under spacetime diffeomorphisms are also gauge invariant order by order in perturbation theory
\cite{MIYO20,MAYO22}.

In the case of interest,~$z$ has no background contribution. Consequently, there is no distinction between gauge invariance and scalar covariance at linear order. At second order, however, only the latter is satisfied in general. The consistency check performed below therefore consists in verifying that the complete second-order expression transforms as a scalar under infinitesimal spacetime diffeomorphisms as in Eq.~\eqref{zscalar}.

\subsection{Covariance of the implicit expression in Eq.~\eqref{zimp}}

Here we show that our expression in Eq.~\eqref{zimp} for the observed pulsar timing modulation~$z$ is a scalar under diffeomorphisms. This implicit expression depends on infinitesimal separations and metric perturbations, and we therefore need to determine how these ingredients transform under an infinitesimal coordinate transformation.

Given that coordinates transform as in Eq.~\eqref{coordtransf}, the transformation of the coordinate intervals~$\Delta x^\mu$ follows directly from taking the difference between transformed coordinates. In the limit of infinitesimal separations, which is the case of our interest here, the result reduces to the standard transformation law of a four-vector evaluated at a fixed spacetime point:
\begin{equation}
    \widetilde{\Delta x}{}^\mu=\frac{\partial \tilde x^\mu}{\partial x^\nu} \Delta x^\nu=\Delta x^\mu+\xi^\mu{}_{,\nu}\Delta x^\nu \,.
\end{equation}
Here and in the following, both sides of the transformation equations are understood to be evaluated at the same physical spacetime point~$\rm p$, which is described by different coordinate values in the two coordinate systems. Thus, the transformed quantity on the left-hand side is evaluated at~$\tilde{x}^\mu_{\rm p}$, while the corresponding quantity on the right-hand side is evaluated at~$x^\mu_{\rm p}$. Unless the arguments are written explicitly, this convention is understood throughout the remainder of the paper.

For convenience, we display the transformations of the temporal and spatial intervals
\begin{equation}\label{transfSep}
    \widetilde{\Delta t}=\Delta t+\dot \xi^0\Delta t+\xi^0{}_{,i}\Delta x^i \,,
    \qquad\qquad\qquad
    \widetilde{\Delta x}{}^i=\Delta x^i+\dot \xi^i\Delta t+\xi^i{}_{,j}\Delta x^j \,.
\end{equation}
These transformations are still exact. Using these results, together with the definition of~$\Theta$ in Eq.~\eqref{Dtrel}, we derive the second-order transformation law governing the shift in the temporal coordinate separations:
\begin{equation}\label{transfDtrel}
    \tilde{\Theta}=\Theta-\dot \xi^0\big\rvert^\emi_\obs-
\frac{(\xi^0_{,i}\Delta x^i)\big\rvert^\emi_\obs}{\widebar{\Delta t}}
+\dot \xi^0\big\rvert^\emi_\obs(\dot \xi^0_\emi-\Theta) \,.
\end{equation}
We remind the reader of the notation~$\rvert^\emi_\obs$ used to denote the difference between the evaluation at the emission~($x^\mu_\emi$) and observation~($x^\mu_\obs$) events associated with the first radio pulse. We now need the coordinate transformation properties of the metric fluctuations. In Appendix~\ref{app:gt}, we have already presented the gauge transformation rules for these quantities; however, here we are interested in their transformation under a coordinate change. So, we start from the well-established gauge transformation relations to derive their coordinate transformation relations.

As discussed in the section summary, these two notions do not coincide beyond linear order. Hence, for quadratic combinations in Eq.~\eqref{zimp}, it is sufficient to use the linear-order gauge transformation rules. By contrast, the second-order gauge transformation of the lapse~$\A$ needs additional contributions for its coordinate transformation. Since the coordinate transformation can be obtained from the gauge transformation by accounting for the shift in the spacetime point, one must relate the evaluation of fields at the same coordinate value in the two charts to their evaluation at the same physical spacetime point. This is achieved by expanding the fields to include the displacement generated by the coordinate transformation, which yields the following additional contribution to the right-hand side of Eq.~\eqref{gtMetric}:
\begin{equation}\label{shift}
    \xi^\lambda \partial_\lam g_{\mu\nu}-2\xi^\lambda\eta_{\rho(\nu}\partial_{\mu)}\partial_\lam\xi^\rho\,,
\end{equation}
where~$g_{\mu\nu}$ is the spacetime metric in Eq.~\eqref{metric} and~$\eta_{\mu\nu}$ is the Minkowski background metric. We thus obtain the coordinate transformation of the lapse at second order
\begin{equation}\label{transfA}
    \tilde \A=\mathcal{A}-  \dot{\xi^0} -  \mathcal{B}^{i}~ \dot{\xi}_{i} -  \frac{1}{2} \dot{\xi}_{i} ~\dot{\xi}^{i}  - 2 \mathcal{A} ~\dot{\xi^0} + \frac{3}{2} \dot{\xi^0}^2   + \dot{\xi}^{i}~ \xi^0_{,i}\,.
\end{equation}
Again, the transformed quantity on the left-hand side is evaluated at~$\tilde{x}^\mu_{\rm p}$, while the corresponding quantity on the right-hand side is evaluated at~$x^\mu_{\rm p}$, where~$\rm p$ denotes an arbitrary spacetime point.

By substituting the first-order transformation rules given in the Appendix~\ref{app:gt}, together with Eqs.~\eqref{transfDtrel} and \eqref{transfA} into Eq.~\eqref{zimp}, we indeed recover the invariance of the observed pulsar timing modulation~$z$ under coordinate transformations, as stated in Eq.~\eqref{zscalar}.

\subsection{Covariance of the explicit expression in Eq.~\eqref{z2}}

We now turn to the explicit expression for the observed pulsar timing modulation~$z$. Having already established from the implicit expression that~$z$ is a scalar under coordinate transformations, it is sufficient to verify that the explicit second-order temporal interval~$\Delta t$ entering~$\Theta^{(2)}$ transforms as required by Eq.~\eqref{transfSep}. Once this is proved, the invariance of the explicit expression in Eq.~\eqref{z2} follows immediately.

Using the relation of~$\Theta$ to~$\Delta t$ in~Eq.~\eqref{Dtrel}, our task amounts to checking the coordinate transformation of~$\Delta t_\obs-\Delta t_\emi$ at second order. The implicit analysis therefore greatly simplifies the consistency check. Rather than verifying the transformation of the full second-order redshift directly, it is enough to establish the correct transformation properties of a small number of intermediate quantities. The second-order temporal interval in Eq.~\eqref{Dt2} consists of a radial contribution  at the boundary points~$\Delta x_{\parallel\emi}-\Delta x_{\parallel\obs}$ together with several additional boundary and integrated terms. To simplify the derivation, we isolate the radial contribution and group all the remaining terms, which is equivalent to the combination~$-(\Delta t+\Delta x_\parallel)\big|^\emi_\obs$. First, we show that the explicit second-order expression for the radial separation in Eq.~\eqref{Dxpara2} transforms as the projection of Eq.~\eqref{transfSep} along the background direction~$\bar n^i$. We will then demonstrate that the remaining contributions in Eq.~\eqref{Dt2} transform exactly the way the combination~$(\Delta t+\Delta x_\parallel)\big|^\emi_\obs$ transforms according to Eq.~\eqref{transfSep} as
\begin{equation}\label{expT}
    \left(\widetilde{\Delta t}+\widetilde{\Delta x}_\pa\right)\Big\rvert^\emi_\obs=\Big(\Delta t+\Delta x_\pa+\Delta x^\mu(\partial_\mu\Xi)\Big) \Big\rvert^\emi_\obs\,,
\end{equation}
where we introduced the quantity~$\Xi\equiv\xi^0+\xi_\parallel$ as a convenient combination of the components of the gauge vector field.

We begin with the first task, namely, verifying the transformation of the second-order radial separation. Starting from Eq.~\eqref{Dxpara2}, we decompose this quantity into the sum of a part that is linear in the perturbations and the other part that is quadratic in the perturbations. The linear term is straightforward to analyze. Since the four-velocity is a four-vector, its spatial components transform accordingly under a coordinate transformation (see Eq.~\eqref{gtU}). It follows that the linear contribution transforms as
\begin{equation}
    \widebar{\Delta t} \, ~\tilde \UU_\parallel=\widebar{\Delta t}
~\Big[\UU_\parallel+\dot\xi_\parallel(1-\A)+\xi_{\parallel,j}\UU^j\Big] \,.
\end{equation}
For the quadratic contribution, it is sufficient to employ the first-order gauge-transformation rules collected in Appendix~\ref{app:gt}. Using these results, we obtain
\begin{equation}
    \left(\widetilde{\Delta t}{}^{(1)}+\widebar{\Delta t}~\tilde\A \right) \tilde\UU_\parallel=\left(\Delta t^{(1)}+\widebar{\Delta t}~\A\right) \UU_\parallel+\dot\xi_\pa\left(\Delta t^{(1)}+\widebar{\Delta t}~\A\right)\,.
\end{equation}
Summing the two expressions, we precisely recover the projection of the transformation law in Eq.~\eqref{transfSep}. This confirms that the explicit second-order equation for the radial separation satisfies the transformation properties required by the implicit analysis.

The second (and the last) task is to verify explicitly the transformation law in Eq.~\eqref{expT}. To this end, we consider all the remaining contributions in Eq.~\eqref{Dt2} other than the radial displacement, and determine their transformation under coordinate transformations. To simplify the presentation, we introduce the notation $\delta_\xi f \equiv \tilde f - f$. For the second-order quantities considered below, no distinction is required between coordinate and gauge transformations, since the difference between evaluating~$\widetilde f$ and~$f$ at the same spacetime point or at the same coordinate value would generate only third-order corrections. Consequently,~$\delta_\xi f$ can be computed directly from the second-order gauge-transformation rules presented in Appendix~\ref{app:gt}. Before proceeding, let us note that the affine parameter is unaffected by spacetime diffeomorphisms in the Minkowski background. As a result, integrations along the null geodesics remain unchanged under coordinate transformations: only the integrands, being spacetime fields, acquire nontrivial transformations.

We begin with the contribution arising from the transformation of~$\Phi$. Substituting the gauge transformation rules for the metric perturbations and the photon wave-vector perturbations in Appendix~\ref{app:gt} into Eqs.~\eqref{deltax} and \eqref{Phi2}, we obtain
\begin{equation}
    \delta_\xi\left(\dot\Phi\rvert_{\bar x_\lam}+{\delta x}{}^\mu_\lam\partial_\mu\dot{\Phi}\right)=\frac{d}{d\lambda}\Big(\dot\Xi\rvert_{\bar x_\lam}+\delta x^\mu\partial_\mu\dot\Xi\Big)+ (\dot{\delta\nu}\,,-  \dot{\delta n}{}^{i})^\mu \partial_\mu\Xi-\dot\xi^\mu\Big(\partial_\mu\Phi+\frac{d}{d\lambda}\partial_\mu\Xi\Big)
    \,.
\end{equation}
Here~$d/d\lambda$ denotes the derivative along the null geodesic. Since the quantity on which it acts is already second order, it is sufficient to evaluate this derivative on the background trajectory, yielding the expression $d/d\lambda=\partial_t-\partial_\parallel$. The same result could alternatively be derived without expanding~$\Phi$ around the background path. In that case, one must use the second-order coordinate transformation of~$\Phi$, which differs from its gauge transformation. However, this coordinate transformation can be conveniently obtained from the gauge transformation by accounting for the shift in the spacetime point, as discussed in Eq.~\eqref{shift}. In this formulation, the derivative along the path is evaluated on the perturbed null geodesic, rather than on the background trajectory, as in the approach presented here. Accounting for this difference leads to the same result.

The remaining contributions are less subtle to handle than the one discussed above. First,  the contribution involving the mismatch in the affine parameter~$\de\lam_e$ transforms simply
\begin{equation}
    \delta_\xi\left({\de\lam}_\emi \Phi_\emi\right)=\de\lam_\emi~(\dot\Xi-\Xi_{,\parallel})_\emi\,,
\end{equation}
showing that~$\de\lambda_\emi$ is gauge invariant. This is merely a coincidence of the perturbative construction and should not be interpreted as implying that~$\de\lambda_\emi$ is an observable quantity. Second, we can use direct substitutions of the gauge transformation rules in Appendix~\ref{app:gt} to show that the contributions involving integrals of the photon wave-vector fluctuations 
transform as
\begin{align}
    \de_\xi \bigg(\int_0^{\lam_\emi} d\lam~\Big(\dot\Phi\int_0^\lam d\hat\lam\,\dot{\de\nu}\Big)\bigg)&=-\dot\xi^0_\obs\int_0^{\lam_\emi}d\lam~\dot\Phi+\int_0^{\lam_\emi}d\lam~\Big(\dot\xi^0~\dot\Phi-\dot\Xi~\dot{\de\nu}+\frac{d\dot\Xi}{d\lam}~\dot\xi^0\Big)
    \nonumber\\&\quad
    -\dot\Xi\Big\rvert^\emi_\obs~\dot\xi^0_\obs+\dot\Xi\Big\rvert_\emi\int_0^{\lam_\emi}d\lam~\dot{\de\nu} \,,
    \\
    \de_\xi \bigg[\int_0^{\lam_\emi} d\lam~\Big(\Phi_{,i}\int_0^\lam d\hat\lam\,(\dot{\de n}{}^i+\eta^i)\Big)\bigg]&=\dot\xi^i_\obs\int_0^{\lam_\emi}d\lam~\Phi_{,i}+\int_0^{\lam_\emi}d\lam~\Big(-\dot\xi^i~\Phi_{,i}-\Xi_{,i}~\dot{\de n}{}^i-\frac{d\Xi_{,i}}{d\lam}~\dot\xi^i\Big)
    \nonumber\\&\quad
    +\Xi_{,i}\Big\rvert^\emi_\obs~\dot\xi^i_\obs+\Xi_{,i}\Big\rvert_\emi\int_0^{\lam_\emi}d\lam~\dot{\de n}{}^i +\eta^i\int_0^{\lam_\emi}d\lam~\lam~\frac{d\Xi_{,i}}{d\lam}\,,
\end{align}
where we have used the fact that~$\eta^i$ is unaffected by spacetime diffeomorphisms. Indeed, it is defined in terms of the observed line-of-sight directions that belong to the local Lorentz frame of the observer, and is therefore independent of both spacetime coordinate transformations and the integration along the photon path. We now derive the gauge transformation of the remaining contribution proportional to the line-of-sight variation~$\eta^i$, which arises from the explicit dependence of~$\Phi$ on the photon-direction perturbation~$\delta n^i$
\begin{equation}
    \de_\xi \bigg(\int _{0}^{\lambda_\emi} d\lambda ~\eta^i\left( \BB_i+2\CC_{\parallel i}+\delta n_i \right)\bigg)=-\eta^i\int _{0}^{\lambda_\emi} d\lambda ~\Xi_{,i}~.
\end{equation}
Finally, the contributions involving infinitesimal coordinate separations at the observation events transform as
\begin{align}
    \de_\xi \Big(\Delta t^{(1)}_\obs \int_0^{\lam_\emi}d\lam~\dot\Phi\Big)&=(\Delta t^{(1)}_\obs+\widebar{\Delta t}~\dot\xi^0_\obs)\dot\Xi\big\rvert^\emi_\obs+\widebar{\Delta t}~\dot\xi^0_\obs\int_0^{\lam_\emi}d\lam~\dot\Phi \,, \\
    \de_\xi \Big(\Delta x^i_\obs \int_0^{\lam_\emi}d\lam~\Phi_{,i}\Big)&=(\Delta x^i_\obs+\widebar{\Delta t}~\dot\xi^i_\obs)\Xi_{,i}\big\rvert^\emi_\obs+\widebar{\Delta t}~\dot\xi^i_\obs\int_0^{\lam_\emi}d\lam~\Phi_{,i} \,.
\end{align}

Combining all the transformation rules derived above, we obtain the transformation rule of the remaining terms in Eq.~\eqref{Dt2} other than the radial component:
\begin{multline}   
\de_\xi\left[(\Delta t+\Delta x_\parallel)\big|^\emi_\obs\right]=\widebar{\Delta t }\,\Big[\dot\Xi(\bar x_\emi)-\dot\Xi(\bar x_\obs)\Big]+\widebar{\Delta t}\left[\delta x^\mu~\partial_\mu\dot\Xi\right]\Big\rvert^\emi_\obs-(\Delta t^{(1)}_\obs\,,\Delta x_\obs^i)^\mu(\partial_\mu\Xi)\Big\rvert_\obs
    \\
+\de\lam_\emi~(\dot\Xi-\Xi_{,\parallel})\big\rvert_\emi+(\partial_\mu\Xi)\big\rvert_\emi\left[(\Delta t^{(1)}_\obs\,,\Delta x_\obs^i)^\mu+\widebar{\Delta t}\int_0^{\lam_\emi}d\lam~(\dot{\delta\nu}\,,-\dot{\de n}{}^i)^\mu\right]
    -\widebar{\Delta t}\,\eta^i \lam_\emi~\Xi_{,i}\big\rvert_\emi
     \,.
\end{multline}
Despite its complicated appearance, these terms add up to be the transformation rule we anticipated in Eq.~\eqref{expT}. The first two square brackets on the right-hand side amount to $\widebar{\Delta t}~\dot\Xi|^\emi_\obs$. The second line is $(\partial_\mu\Xi)\big\rvert_\emi(\Delta t^{(1)}_\emi\,,\Delta x_\emi^i)^\mu$, once we note that the term involving the shift in the affine parameter can be expressed as $\de\lam_\emi(\dot\Xi-\Xi_{,\parallel})\big\rvert_\emi=\de\lam_\emi(\partial_\mu\Xi)\big\rvert_\emi(1\,,-\bar n^i)^\mu$ and recast $\eta^i \lam_\emi\Xi_{,i}\big\rvert_\emi=(\partial_\mu\Xi)\big\rvert_\emi\left(0,\int_0^{\lam_\emi}d\lam~\eta^i \right)^\mu$. With the remaining third term in the first line, the second line can be combined to yield $(\partial_\mu\Xi)(\Delta t^{(1)}\,,\Delta x^i)^\mu\big|^\emi_\obs$. Thus, we explicitly recovered the transformation expected for the combination~$(\Delta t+\Delta x_\parallel)\big|^\emi_\obs$.

\section{Summary and Discussion}
\label{sec:disc}
In this work, we have developed a fully relativistic description of the observed pulsar timing modulation and derived its complete expression to second order in perturbations around a Minkowski background, without fixing a gauge and retaining the full scalar, vector, and tensor content of the metric perturbations. We first formulated the observable at the fully nonlinear level, without performing any perturbative expansion of the spacetime geometry. The pulsar timing modulation is defined from the proper time intervals between successive emission and observation events in Eq.~\eqref{zdef}, and we then consider the limit in which the emission interval tends to zero. Importantly, this limit concerns only the separation between successive pulses and is independent of the perturbative expansion of the spacetime geometry. Combining the exact relations along the pulsar and observer worldlines with the propagation of neighboring null geodesics leads to the fully nonlinear expression in Eq.~\eqref{Dt}, which provides the basis for the perturbative calculation.

Specializing the fully nonlinear formalism to first order, we obtain the pulsar timing modulation in Eq.~\eqref{z1}. Upon fixing the transverse-traceless gauge, our expression recovers the standard result of the PTA literature \cite{DETWE79,ANBAET09}. Conventional derivations typically impose this gauge from the outset, whereas our general expression makes explicit the physical origin of the additional terms. In particular, the lapse contribution arises from the conversion between coordinate time and proper time intervals, as shown in Eq.~\eqref{lapse}, while the velocity terms describe the Doppler contribution associated with the pulsar and observer motion. Their combination with the metric perturbations makes explicit how coordinate independence is realized before any gauge fixing is made. The central part of this work is the second-order derivation presented in Sec.~\ref{sec:second}. Here the nonlinear formulation is expanded consistently to second order, where several effects absent at first order must be retained. These include the displacement of the emission and observation events, Eqs.~\eqref{Dxpara2}, \eqref{Dobs}, and \eqref{delam1}, the perturbation of the photon path, Eqs.~\eqref{deltax} and \eqref{PhiTWO}, and the change in the observed line of sight between successive pulses, Eqs.~\eqref{gamma'gamma} and \eqref{Phiexp}. Their consistent combination, together with the second-order metric and velocity perturbations, yields the complete second-order observed timing modulation in Eq.~\eqref{z2}.

The completeness of Eq.~\eqref{z2} is then tested in two independent ways. First, as the pulsar timing modulation is an observable quantity, the final result must be independent of the coordinates used to describe the spacetime. In Sec.~\ref{sec:gauge}, we verify this explicitly by deriving the transformation of the quantities entering the timing modulation and showing that their combination obeys the scalar transformation law in Eq.~\eqref{zscalar}. This provides a stringent test at second order, where the individual perturbative quantities are coordinate dependent although the observable itself is not. A separate test follows from the exact nonlinear relation established in Ref.~\cite{MAGIYOO26}. In the limit where the emission interval tends to zero, the pulsar timing modulation must coincide with the observed redshift defined along a single null geodesic. In Appendix~\ref{app:z}, we specialize the fully relativistic second-order observed redshift derived in Ref.~\cite{MAYO22} to the Minkowski background and compare it directly with Eq.~\eqref{z2}. The two expressions agree exactly, providing a strong check that all contributions required at second order have been consistently accounted for.

A natural application of the second-order result is the contribution of scalar-induced gravitational waves to the observed pulsar timing modulation. At second order in perturbation theory, terms quadratic in first-order scalar perturbations source tensor perturbations \cite{MOHAMA04,BASTET07,ANCLWA07}. Such scalar-induced gravitational waves have been investigated as a possible origin of the signals reported by PTA collaborations \cite{NANO23new,EPTA23d,CAHEET23,INKOTE24,FRIOET23}.

The theoretical description of these waves requires distinguishing the cosmological tensor perturbations from their observable effects. The induced tensor power spectrum is usually computed in a specified gauge \cite{BASTET07,ANCLWA07,PISA20}, but tensor perturbations are gauge dependent beyond linear order \cite{HWJENO17,DEFRET20,INTE20}. Various recent approaches have investigated gauge-invariant formulations, compared the induced tensor spectrum between different gauges, or identified conditions under which the resulting spectrum becomes approximately gauge independent \cite{DOSA18,DEFRET20,INTE20,ALGOLU21,DOSA21}. While these approaches provide useful information about the gauge dependence of the induced tensor perturbations, they do not resolve the underlying problem. The induced tensor perturbation is not itself an observable and therefore cannot in general be assigned a direct physical meaning. A physical prediction should instead be obtained by computing the corresponding observable directly, in which all gauge-dependent contributions combine into a gauge-independent result.

This is precisely the approach taken in the present work. Rather than identifying a particular metric component with the PTA signal, we have derived the complete theoretical description of the pulsar timing modulation measured by the observer to second order, without fixing a gauge and while retaining the full scalar, vector, and tensor content of the metric perturbations, together with the complete light-propagation contributions. The same time delay observable was recently studied by Dom\`enech, Pi, and Wang \cite{DOPIWA26}, who considered geodesic observers exchanging electromagnetic signals and derived the corresponding time delay using an approach different from ours. At second order, their calculation focuses on the contributions that give a quadrupolar signature in the time delay, including both scalar and tensor contributions, and they express the result in terms of gauge-invariant quantities, making the gauge invariance manifest. Our calculation instead derives the complete second-order PTA timing observable, without restricting to the quadrupolar sector and retaining all perturbative contributions.

A crucial aspect of our derivation is that the complete second-order result is subject to two nontrivial and independent consistency checks. First, we verify the coordinate independence explicitly by substituting the full second-order transformation laws into the complete timing modulation in Eq.~\eqref{z2} and demonstrating the cancellation of all coordinate-dependent contributions. Second, in the limit of a vanishing emission interval, we show explicitly that the timing modulation obtained from two neighboring radio pulses [Eq.~\eqref{z2}] reduces to the observed redshift computed along a single null geodesic [Eq.~\eqref{zsingle}], thereby realizing at second order the equivalence established in \cite{MAGIYOO26}. Together, these provide particularly powerful consistency checks of the full second-order calculation, since subtle propagation and boundary contributions can easily be missed, while the successful completion of both checks gives confidence that all required contributions to the complete timing modulation have been included consistently.

The scalar-induced gravitational-wave signal in PTA observations can therefore be studied by combining our observable response with the second-order evolution generated from scalar initial conditions. This makes it possible to determine consistently how the different second-order contributions enter the timing residuals and the angular correlations between pulsars. In particular, the pulsar cross-correlation may receive contributions absent at linear order, introducing additional angular dependence and sensitivity to higher-order statistics of the underlying perturbations, especially for non-Gaussian sources. We leave these applications for future work.

\acknowledgments
We acknowledge useful discussions with Guillem Dom\`enech. We are particularly grateful to Julian Adamek and Chris Hirata for valuable comments and discussions during the early development of this work. MM is supported by the Institute for Basic Science under the project code IBS-R018-D3.

\appendix

\section{Linear-Order Expression for the Photon Wave-Vector}
\label{app:k}
In this appendix we derive the explicit expressions for the first-order photon wave-vector perturbations used in the main text. Integrating the null geodesic equation along the background light path up to an arbitrary point gives the following form of Eq.~\eqref{geoNL}:
\begin{align}
    \delta\nu(\bar x_{\lambda}) = \delta\nu_\obs - \int_{\lambda_\obs}^{\lambda}d\hat\lambda\, \Gamma^0{}_{\mu\nu}(\bar x_{\hat\lambda}) \bar k^\mu\bar k^\nu \,,
    \qquad\qquad\quad
    \delta n^i(\bar x_{\lambda}) = \delta n^i_\obs + \int_{\lambda_\obs}^{\lambda}d\hat\lambda\, \Gamma^{i}{}_{\mu\nu}(\bar x_{\hat\lambda}) \bar k^\mu\bar k^\nu \,.
\end{align}
Since the background Christoffel symbols vanish in Minkowski spacetime in Cartesian coordinates, the wave vector appearing in the integrands above is the background one. According to the decomposition introduced in Eq.~\eqref{ksplit}, it reads
\begin{equation}
    \bar k^\mu=\left(1,\,-\bar n^i \right)^\mu\,.
\end{equation}
The first-order Christoffel symbols associated with the metric in Eq.~\eqref{metric} are
\begin{align}
    \Gamma^0{}_{00} &= \dot{\mathcal A}\,,&
    \Gamma^0{}_{0i} &= \mathcal A_{,i}\,,&
    \Gamma^0{}_{ij} &= \mathcal B_{(i,j)} +\dot{\mathcal C}_{ij}\,,
    \nonumber\\
    \Gamma^i{}_{00} &= \mathcal A^{,i}-\dot{\mathcal B}^{i}\,,&
    \Gamma^i{}_{0j} &= \dot{\mathcal C}^{i}{}_{j}-\mathcal B^{[i}{}_{,j]}\,,&
    \Gamma^i{}_{jk} &= 2\mathcal C^i{}_{(j,k)}-\mathcal C_{jk}{}^{,i}\,.
\end{align}
Substituting these expressions into the integrated geodesic equations above, we obtain the first-order wave-vector fluctuations up to integration constants
(see, e.g., \cite{YOO14a})
\begin{align}\label{k1}
    \delta\nu(\bar x_\lambda) &= \delta\nu_\obs +\left(\mathcal B_\parallel-2\mathcal A  \right)\big\rvert_{\obs}^{\lambda}
     +\int_{0}^{\lambda} d\hat\lambda\,\left(\dot\A-\dot\BB_\parallel -\dot\CC_{\parallel\parallel}\right)\,,
    \\\label{k11}
    \delta n^i(\bar x_\lambda) &= \delta n^i_\obs -\left(\mathcal B^i+2\mathcal C^i_\parallel  \right)\Big\rvert_{\obs}^{\lambda} +\int_{0}^{\lambda}d\hat\lambda\,  \left(\A^{,i}-\BB_\parallel{}^{,i} -\CC_{\parallel\parallel}{}^{,i}\right)\,.
\end{align}
The integration constants~$\delta\nu_\obs$ and~$\delta n^i_\obs$ are fixed by the boundary conditions imposed at the observer position, according to Eq.~\eqref{boundary}. To implement these conditions, we introduce an orthonormal tetrad~$e_A^\mu$ at the observer, satisfying
\begin{equation}
    g_{\mu\nu}e_A^\mu e_B^\nu=\eta_{AB}\,,
    \qquad\qquad
    e^A_\mu e_B^\mu=\delta^A_B \,.
\end{equation}
Here and in the following, $A,B,C,\ldots=0,1,2,3$ and $I,J,K,\ldots=1,2,3$ denote internal Lorentz indices, which should be distinguished from spacetime indices.

We fix the boost part of the local Lorentz freedom by aligning the timelike tetrad vector with the observer four-velocity,~$e_0^\mu\equiv u^\mu$. The spatial triad is then constrained by the orthonormality conditions. At linear order, it can be written as \cite{YOGRET18,MIYO20}
\begin{equation}
    e^\mu_I=\Big( \delta^i_I\UU_i-\delta^i_I\mathcal{B}_i\,,\, \delta^i_I-\delta^i{}_J (   \mathrm S^J{}_I + \mathrm A^J{}_I )\Big)^\mu\,, \qquad \qquad
\mathrm S_{IJ} = \delta_I^i\delta_J^j \mathcal C_{ij}\,,\qquad
    \mathrm A_{IJ}=-\mathrm A_{JI}\,.
\end{equation}
The symmetric part~$\mathrm{S}_{IJ}$ is fixed by the spatial metric perturbation, whereas the antisymmetric part~$\mathrm A_{IJ}$ is not determined by the metric alone. Its form depends on the remaining internal spatial rotation of the tetrad. A convenient parameterization, consistent with its transformation properties under spacetime diffeomorphisms, is
\begin{equation}
    \delta^i_J \mathrm A^J{}_I =  C^{[i}{}_{,j]}\delta_I^j +\epsilon^i{}_{IJ}\Omega^J \,.
\end{equation}
Here~$C_i$ is the transverse vector mode entering the spatial metric perturbation~$\mathcal C_{ij}$ in the scalar-vector-tensor (SVT) decomposition \cite{BARDE80}, while~$\Omega^I$ parameterizes the orientation of the spatial triad with respect to the background spatial coordinates. Once this residual internal rotation is specified, the tetrad is fully fixed.

We thus obtain the boundary conditions \cite{YOGRET18}
\begin{align}\label{boundarys1st}
    \delta \nu_\obs = -\left(\A + \UU_{\parallel}-\BB_{\parallel}\right)_\obs \,,
    \qquad\qquad\qquad
    (\delta n_i)_\obs= -\bar n_i+\left(\delta_i^I\hat n_I-\UU_i- \CC_{i\parallel}- C_{[i,\parallel]}- (\bm{\hat n}\times\bm\Omega)_i\right)_\obs \,.
\end{align}
These boundary conditions, together with Eqs.~\eqref{k1} and \eqref{k11}, fully determine the linear photon wave-vector fluctuations along the light path.

The same construction applies at the observation event of the second radio pulse. In that case, the metric and matter perturbations are generally different, as they are evaluated at a different spacetime point. Likewise, the residual internal rotation of the tetrad may be specified independently of that at the first observation event. Finally, since the second pulse is received at a different event on the observer worldline, its observed line-of-sight direction is also, in general, different.

\section{Second-Order Gauge Transformations}
\label{app:gt}

We collect in this appendix the gauge transformation rules employed throughout this work. The purpose is only to fix notation and conventions for the transformations used in the main text; complete derivations and further details can be found in \cite{MAYO22}, whose conventions we follow closely.

It is useful to distinguish tensorial coordinate transformations from perturbative gauge transformations. Throughout this work, quantities expressed in the transformed coordinate system are denoted by a tilde. The two coordinate systems are related by the infinitesimal transformation in Eq.~\eqref{coordtransf}. A tensorial transformation relates the components of a given geometrical object at the same physical spacetime point, described in two different coordinate systems. The physical point is unchanged; only its coordinate labels and the corresponding tensor components are transformed. By contrast, a gauge transformation compares perturbations evaluated at the same coordinate values in the two coordinate systems. These identical coordinate values do not correspond to the same physical spacetime point, because changing the gauge changes the correspondence between the background spacetime and the perturbed spacetime. The gauge transformation rules are therefore obtained by expanding the tensorial transformation laws and then evaluating the transformed and untransformed quantities at identical coordinate values.

This distinction becomes particularly important beyond linear order. A gauge transformation differs from a coordinate transformation in the physical point at which the transformed and untransformed quantities are compared, and relating the two therefore requires shifting the evaluation point. At linear order, this shift acts only on the background quantity, so that for a constant background the two transformations give the same result. Beyond linear order, however, the shift also acts on lower-order perturbations, and the two transformations generally differ even when the background is constant.

To derive the gauge transformation of the metric components we start from the tensorial transformation
\begin{equation}
\tilde g_{\mu\nu}
=
\frac{\partial x^\rho}{\partial \tilde x^\mu}
\frac{\partial x^\tau}{\partial \tilde x^\nu}~~
g_{\rho\tau} \,.
\end{equation}
Expanding this relation perturbatively up to second order, and evaluating the transformed and untransformed tensors at the same coordinate values, we obtain the following gauge transformations of the metric perturbations
(see, e.g., \cite{ACBAET03,NOHW04,MAWA09,BAKOET04}):
\begin{align}\label{gtMetric}
    \tilde \A\Big\rvert_{\tilde x=x}&=\mathcal{A}-  \dot{\xi^0} -  \mathcal{B}^{i} \dot{\xi}_{i} -  \frac{1}{2} \dot{\xi}_{i} \dot{\xi}^{i} -  \dot{\mathcal{A}} \xi^0 - 2 \mathcal{A} \dot{\xi^0} + \frac{3}{2} \dot{\xi^0}^2 + \xi^0 \ddot{\xi^0}  -  \xi^{i} \mathcal{A}_{,i} + \dot{\xi}^{i} \xi^0_{,i} + \xi^{i} \dot{\xi^0}_{,i}\,,
    \nonumber\\
    \tilde \BB_i\Big\rvert_{\tilde x=x}&=\mathcal{B}_{i} -  \xi^0_{,i}+ \dot{\xi}_{i} + 2 \mathcal{C}_{ij} \dot{\xi}^{j}  -  \dot{\mathcal{B}}_{i} \xi^0 -  \ddot{\xi}_{i} \xi^0 -  \mathcal{B}_{i} \dot{\xi^0} -  \dot{\xi}_{i} \dot{\xi^0} -  \mathcal{B}^{j} \xi_{j,i} - 2 \dot{\xi}^{j} \xi_{(i,j)} - 2 \mathcal{A} \xi^0_{,i} 
    \nonumber\\&\quad
    + 2 \dot{\xi^0} \xi^0_{,i} +\xi^0 \dot{\xi^0}_{,i} -  \xi^{j} \mathcal{B}_{i,j}  -  \xi^{j} \dot{\xi}_{i,j} + \xi^{j}{}_{,i} \xi^0_{,j} + \xi^{j} \xi^0_{,ji}\,,
    \nonumber\\
    \tilde\CC_{ij}\Big\rvert_{\tilde x=x}&=\mathcal{C}_{ij}-  \xi_{(i,j)} -  \dot{\mathcal{C}}_{ij} \xi^0 -  2\mathcal{C}_{k(i} \xi^{k}{}_{,j)} +  \xi^0 \dot{\xi}_{(i,j)}  +  \mathcal{B}_{(i} \xi^0_{,j)} + \dot{\xi}_{(i} \xi^0_{,j)} 
    + \frac{1}{2} \xi^{k}{}_{,i} \xi_{k,j}  -  \frac{1}{2} \xi^0_{,i} \xi^0_{,j}
    \nonumber\\&\quad
    -  \xi^{k} \mathcal{C}_{ij,k} +  \xi^{k} \xi_{(i,j)k} +  \xi^{k}{}_{,(i} \xi_{j),k}\,,
\end{align}
where the transformed and untransformed quantities are both evaluated at the same coordinate value~$x^\mu$.

For the photon wave vector, we start from its tensorial four-vector transformation law,
\begin{equation}
\tilde{k}^\mu
=
\frac{\partial \tilde x^\mu}{\partial x^\nu} k^\nu
=
k^\mu+\xi^\mu{}_{,\nu}k^\nu \,.
\end{equation}
Using the decomposition introduced in Eq.~\eqref{ksplit} and expanding to first order, we obtain the corresponding gauge transformation of the wave-vector fluctuations:
\begin{equation}\label{GTnu}
\widetilde{\delta\nu}
=
\delta\nu +\frac{d\xi^0}{d\lambda} \,,
\qquad\qquad\qquad
\widetilde{\delta n}{}^i
=
\delta n^i -\frac{d\xi^i}{d\lambda} \,,
\end{equation}
where $\tfrac{d}{d\lam}\equiv \partial_t-\partial_\parallel$ denotes the derivative along the background light path. Integrating the transformed photon trajectory according to Eq.~\eqref{deltax} then gives the corresponding transformation of the distortions to the background path,
\begin{equation}\label{GTdX}
\widetilde{\delta x}{}^\mu_\lambda
=
\delta x^\mu_\lambda +\xi^\mu_\lam \,,
\qquad\qquad\qquad
\widetilde{\delta x}{}^\mu_\obs
=
\delta x^\mu_\obs +\xi^\mu_\obs \,.
\end{equation}
Note, however, that Eqs.~\eqref{GTnu} and~\eqref{GTdX} involve more linear-order contributions if the background is the Robertson-Walker metric, since the affine parameters corresponding to the same physical points are different in two different coordinates \cite{YOGRET18,MAYO22}.

The observer four-velocity is also a spacetime vector field and therefore obeys 
\begin{equation}
\tilde{u}^\mu
=
\frac{\partial \tilde x^\mu}{\partial x^\nu} u^\nu
=
u^\mu+\xi^\mu{}_{,\nu}u^\nu \,.
\end{equation}
According to the decomposition introduced in Eq.~\eqref{splitu}, the temporal component is determined by the normalization condition~$u_\mu u^\mu=-1$ in Eq.~\eqref{normu}. Its transformation therefore follows directly from the corresponding transformation of the metric perturbations given in Eq.~\eqref{gtMetric}. Instead, the spatial component, corresponding to the peculiar velocity, transforms up to second order as
\begin{equation}\label{gtU}
\tilde{\UU}^i
=
\UU^i+\dot\xi^i(1-\A)+\xi^i{}_{,j}\UU^j \,.
\end{equation}
In the main text, we make use of both the second-order coordinate transformation above and the corresponding first-order gauge transformation. The latter is obtained by retaining only the linear terms in the transformation law, since coordinate and gauge transformations coincide at first order.

\section{Equivalence to the Observed Redshift}
\label{app:z}
Ref.~\cite{MAGIYOO26}
established a fully nonlinear equivalence that in the vanishing-period limit, the observed pulsar timing modulation~$z$ defined from the ratio of the proper-time intervals between two successive emissions and observations  is exactly equivalent to the observed redshift along a single light path. Note that the former is the interval of two times of light signal arrival along two different geodesic paths regardless of their photon wavelengths, while the latter is the change of physical wavelength along a single geodesic path. Since these two phenomena are physically distinct, the proof is indeed valid only in the limit of the vanishing proper-time interval (which we adopted throughout the manuscript). The explicit second-order expression derived in Eq.~\eqref{z2} must therefore be equivalent to the second-order expression for the observed redshift. In this appendix, we explicitly demonstrate the equivalence at second order in perturbations. The observed redshift is defined as the ratio of the rest-frame frequency~$\omega_\emi$ to the observed frequency~$\omega_\obs$ as
\begin{equation}
    1+\hat z=\frac{\omega_\emi}{\omega_\obs}\,,
\end{equation}
where the hat distinguishes the single-light-path redshift from the two-pulse observed redshift introduced in the main text. The rest-frame frequency at the emission or the observation can be obtained as
\begin{equation}\label{freq}
    \omega=-u^\mu k_\mu\,.
\end{equation}

The observed redshift associated with a single null geodesic \cite{SAWO67} has been studied extensively in the literature in an FLRW background. In what follows, we adopt the expression derived in our previous work \cite{MAYO22} and take consistently the Minkowski limit ($a\equiv1$) of the Robertson-Walker metric. The rest-frame frequency at second order then reads
\begin{equation}\label{dhnu}
    \omega=1+ \de\nu+\A+\UU_\parallel-\BB_\parallel-\frac12\A^2+\frac12\UU_i~\UU^i+\A~\de\nu+\delta n^i(\UU_i-\BB_i )+\A~\BB_\parallel+2\CC_{i\parallel} ~\UU^{i}
    \,.
\end{equation}

Since the rest-frame frequency in Eq.~\eqref{freq} is a scalar, it may equivalently be evaluated in the local Lorentz frame defined by the observer tetrad. Our choice of affine parameterization fixes the photon wave vector at the observer to be $k^A_\obs=(1,-\bm{\hat n})^A$, implying that the observed rest-frame frequency is normalized to unity,
\begin{equation}\label{freq1}
    \omega_\obs=-\left(e^\mu_0e^A_\mu k_A\right)_\obs=-\left(\delta^A_0 k_A
\right)_\obs=1\,.
\end{equation}
Note that our metric convention is mostly plus. It follows immediately that the combination of the perturbations in Eq.~\eqref{dhnu} vanishes at the observer position. Thus,
\begin{equation}\label{zsingle}
    \hat z=\Big( \de\nu+\A+\UU_\parallel-\BB_\parallel-\frac12\A^2+\frac12\UU_i~\UU^i+\A~\de\nu+\delta n^i(\UU_i-\BB_i )+\A~\BB_\parallel+2\CC_{i\parallel} ~\UU^{i} \Big)_\emi\,.
\end{equation}

To establish the equivalence between Eq.~\eqref{z2} and Eq.~\eqref{zsingle}, it is convenient to define the following linear-order combinations:
\begin{equation}\label{simp1}
    E\equiv\de\nu-\BB_\parallel+2\A\,,\qquad\quad N_i\equiv\delta n_i+\BB_i+2\CC_{i\parallel}\,,\qquad\quad I_\nu (\lam)\equiv\int_0^\lam d\lam~\dot{\de\nu}\,,\qquad\quad I^i_n (\lam)\equiv\int_0^\lam d\lam~\dot{\de n}{}^i\,,
\end{equation}
which, from Eqs.~\eqref{null}, \eqref{PHI}, \eqref{k1}, and~\eqref{k11}, satisfy
\begin{equation}\label{simp2}
    \frac{d}{d\lam}E=-\dot{\Phi}\,,\qquad\qquad \qquad
\frac{d}{d\lam}N_i=-\Phi_{,i}\,,\qquad\qquad \qquad
E_\obs=\A_\obs-\UU_{\parallel \obs}\,.
\end{equation}

We begin by rewriting Eq.~\eqref{z2} in terms of the quantities just defined, obtaining the equivalent expression
\begin{multline}
z
= \left[\UU_\parallel-\A+\UU_\parallel\A+\frac12(\A^2+\UU_i\UU^i)-\BB_i\UU^i\right]\bigg\rvert^{\emi}_{\obs}+\A_\emi\,\A\rvert^\emi_\obs-\int _{0}^{\lambda_\emi} d\lambda ~\left[\dot\Phi+\delta x^\mu_{\lambda} \,(\partial_\mu \dot\Phi)\right]\Big\rvert_{\bar x_{\lambda}}
\\
-\eta{}^i\int_0^{\lam_\emi} d\lam~\Big(\lam\frac{d}{d\lam}N_i+N_i\Big)+\int_0^{\lam_\emi} d\lam~\left(\frac{dE}{d\lam}I_\nu-\frac{dN_i}{d\lam}I^i_n\right)
+\UU^i_\obs N_i\big\rvert^\emi_\obs+
\Theta\left(N_\parallel-\A \right)_\emi
+\Phi_\emi I_\nu(\lam_\emi)
\,.
\end{multline}
Next, we integrate by parts the terms containing~$dE/d\lambda$ and~$dN_i/d\lambda$. Since the integral functions~$I_\nu$ and~$I_n^i$ vanish at the observer position, only boundary terms at the emission survive. Moreover, the integral proportional to~$\eta^i$ reduces to the boundary contribution $\eta^iN_i\big\rvert_\emi d$. The two boundary terms proportional to~$I_\nu(\lam_\emi)$ combine through the identity $E+\Phi=N_\parallel$, so that the remaining endpoint contribution is written in terms of $N_\parallel I_\nu-N_iI_n^i$. Finally, using the relation $I_\nu-I_n^\parallel=-E\big\rvert^\emi_\obs$, we derive
\begin{multline}
z
= \left[\UU_\parallel-\A+\UU_\parallel\A+\frac12(\A^2+\UU_i\UU^i)-\BB_i\UU^i\right]\bigg\rvert^{\emi}_{\obs}+\A_\emi\,\A\rvert^\emi_\obs-\int _{0}^{\lambda_\emi} d\lambda ~\left[\dot\Phi+\delta x^\mu_{\lambda} \,(\partial_\mu \dot\Phi)\right]\Big\rvert_{\bar x_{\lambda}}+\eta{}^iN_{i}\big\rvert_\emi d
\\
-\int_0^{\lam_\emi} d\lam~\left(E\,\dot{\de\nu}-N_{i}\dot{\de n}{}^i\right)
+\UU^i_\obs N_i\big\rvert^\emi_\obs+
\Theta\left(N_\parallel-\A \right)_\emi
-N_\parallel\big\rvert_\emi E\big\rvert^\emi_\obs
+N_\parallel\big\rvert_\emi I^\parallel_n(\lam_\emi)- N_i\big\rvert_\emi I^i_n(\lam_\emi)
\,.
\end{multline}
We now decompose the contraction~$N_iI_n^i$ into components parallel and orthogonal to the background propagation direction~$\bar n^i$. The longitudinal component cancels against the explicit term~$N_\parallel I_n^\parallel$, while the transverse component is eliminated upon substituting the expression for~$\eta^i$ in Eq.~\eqref{eta}. The resulting expression therefore reduces to
\begin{align}
z
&= \left[\UU_\parallel-\A+\UU_\parallel\A+\frac12(\A^2+\UU_i\UU^i)-\BB_i\UU^i\right]\bigg\rvert^{\emi}_{\obs}+\A_\emi\,\A\rvert^\emi_\obs-\int _{0}^{\lambda_\emi} d\lambda ~\left[\dot\Phi+\delta x^\mu_{\lambda} \,(\partial_\mu \dot\Phi)\right]\Big\rvert_{\bar x_{\lambda}}
\nonumber\\&\quad
+N^i_\perp\big\rvert_\emi\UU_{\perp i}\big\rvert^\emi_\obs-\int_0^{\lam_\emi} d\lam~\left(E\,\dot{\de\nu}-N_{i}\dot{\de n}{}^i\right)
+\UU^i_\obs N_i\big\rvert^\emi_\obs+
(\UU_\parallel+E)\big\rvert^\emi_\obs\left(N_\parallel-\A \right)_\emi
-N_\parallel\big\rvert_\emi E\big\rvert^\emi_\obs
\,,
\end{align}
where we have further used Eqs.~\eqref{Dt1} and \eqref{simp2} to write $\Theta^{(1)}=(\UU_\parallel+E)\big\rvert^\emi_\obs$.

Using the covariant geodesic equation for the temporal component of the photon wave vector, $k_0$, the remaining line-of-sight integrals can be converted into endpoint terms. In fact, the temporal geodesic equation yields
\begin{equation}
\frac{d}{d\lambda}
\left(
\de\nu-\BB_\parallel+2\A
+2\A\de\nu-\BB_i\delta n^i
\right)
=
-\left[
\dot\Phi+\delta x^\mu_\lambda(\partial_\mu\dot\Phi)
\right]_{\bar x_\lambda}
-E\dot{\de\nu}
+N_i\dot{\de n}{}^i \,.
\end{equation}
Integrating this relation along the background null geodesic converts the remaining line-of-sight integrals into endpoint contributions. After substituting the resulting boundary expression into the previous equation, the linear terms in the lapse combine, while the quadratic contributions are carried along unchanged. We thus obtain
\begin{align}
    z&= \Big[\de\nu+\A+\UU_\parallel-\BB_\parallel+\UU_\parallel\A+\frac12(\A^2+\UU_i~\UU^i)-\BB_i~\UU^i+2\A~\de\nu-\BB_i \delta n^i\Big]\Big\rvert^{\emi}_{\obs}+\A_\emi\,\A\big\rvert^\emi_\obs
    \nonumber\\&\quad
    +N^i_{\perp\emi}~\UU_{\perp i}\big\rvert^\emi_\obs+\UU^i_\obs ~N_i\big\rvert^\emi_\obs+
    (\UU_\parallel+E)\big\rvert^\emi_\obs\left(N_\parallel-\A \right)_\emi-N_\parallel\big\rvert_\emi E\big\rvert^\emi_\obs
    \,.
\end{align}
Substituting the definitions of $E$ and $N_i$ in Eq.~\eqref{simp1}, together with the unit-frequency condition at the observer in Eq.~\eqref{freq1}, all terms evaluated at the observer cancel exactly. We thus conclude
\begin{align}
z
&= \Big(\de\nu+\A+\UU_\parallel-\BB_\parallel-\frac12\A^2+\frac12\UU_i~\UU^i+\A~\de\nu+\delta n^i(\UU_i-\BB_i )+\A~\BB_\parallel+2\CC_{i\parallel} ~\UU^{i}\Big)_\emi
\,,
\end{align}
which indeed coincides with Eq.~\eqref{zsingle}.

The equality $z=\hat z$ is expected from the nonlinear identity established in Ref.~\cite{MAGIYOO26}. Recovering it explicitly at second order is nevertheless highly nontrivial, requiring the exact cancellation and recombination of all auxiliary quantities introduced by the two-pulse construction. The agreement with the independently derived single-pulse expression provides a strong consistency check of both calculations.

\bibliographystyle{JHEP}
\bibliography{ms.bbl} 

\end{document}